\documentclass[a4paper,11pt]{article}
\pdfoutput=1
\usepackage{jheppub}

\usepackage{epsf}
\usepackage{epsfig}
\usepackage{subfigure}
\usepackage{mathtools}
\usepackage{hhline}
\usepackage{float}
\usepackage{multirow}
\usepackage{nicefrac}
\usepackage{epstopdf}
\usepackage{url}
\usepackage{multicol}
\usepackage{array}
\usepackage[normalem]{ulem}
 
\allowdisplaybreaks

\newcommand{\be}{\begin{eqnarray}}
\newcommand{\ee}{\end{eqnarray}}

\newcommand{\ba}{\begin{array}}
\newcommand{\ea}{\end{array}}
\newcommand{\bee}{\begin{equation}\ba{c}}
\newcommand{\eee}{\ea\end{equation}}

\newcommand{\bi}{\begin{itemize}}
\newcommand{\ei}{\end{itemize}}

\toccontinuoustrue

\title{Signals with six bottom quarks for charged and neutral Higgs bosons}

\author{Radovan Dermisek$^1$,}
\author{Enrico Lunghi$^1$}
\author{Navin McGinnis$^1,^2$}
\author{Seodong Shin$^{3}$}
\affiliation{
$^1$Physics Department, Indiana University, Bloomington, IN 47405, USA \\
$^2$High Energy Physics Division, Argonne National Laboratory, Lemont, IL 60439, USA\\
$^3$Department of Physics, Jeonbuk National University, Jeonju, Jeonbuk 54896, Korea\\ 
}
\emailAdd{dermisek@indiana.edu} 
\emailAdd{elunghi@indiana.edu} 
\emailAdd{sshin@jbnu.ac.kr}
\emailAdd{nmmcginn@indiana.edu}

\abstract{In extensions of two Higgs doublet models with vectorlike quarks, the decays of vectorlike quarks may easily be dominated by cascade decays through charged or neutral Higgs bosons leading to signatures with 6 top or bottom quarks. Since top quark decays also contain bottom quarks, a common signature for many possible decay chains is 6 bottom quarks in the final state. We present a search strategy focusing on this final state and find the mass ranges of vectorlike quarks and Higgs bosons that can be explored at the Large Hadron Collider. Among other results, the sensitivity to the charged and neutral Higgs bosons, extending to about 2 TeV, stands out when compared to models without vectorlike matter.
}

\begin{document}

\maketitle

\section{Introduction}
\label{sec:intro}

Among the simplest extensions of the standard model (SM) is the two Higgs doublet model (2HDM) featuring two neutral Higgs bosons, the CP-even, $H$, and the CP-odd, $A$, (if CP is conserved) and a pair of charged Higgs bosons, $H^\pm$, in addition to the neutral CP-even Higgs boson, $h$, which is experimentally constrained to have properties very close to the SM Higgs boson. However, the typically dominant decay modes of the additional Higgs bosons, $t \bar t$, $b \bar b$, $t\bar b$ and $\bar t b$, in the type-II 2HDM are very challenging at the Large Hadron Collider (LHC) due to huge QCD backgrounds. For example, the limits on $H^+ \to  t \bar b$ currently exist only for small and large $\tan \beta$ (the ratio of vacuum expectation values of two Higgs doublets, that set the Yukawa couplings of Higgs bosons) and the LHC is not yet sensitive to this decay in most of the range, $2 \lesssim \tan \beta \lesssim 35$~\cite{Aaboud:2018cwk, Sirunyan:2019arl}. Although the limits will certainly improve with larger data sets, even with 3 ab$^{-1}$ of integrated luminosity it is not expected that the LHC would constrain the charged Higgs boson significantly in this range.\footnote{Similarly, the neutral Higgs bosons are constrained by $H(A) \to t\bar t$ only at small $\tan \beta$~\cite{Sirunyan:2019wph}, and at large $\tan \beta$ it is actually the subleading  $H(A) \to \tau \bar \tau$ decay mode which leads to stronger limits than  $H(A) \to b\bar b$~\cite{Sirunyan:2018taj, Aad:2019zwb}. This is also expected to be the case with larger data sets~\cite{CidVidal:2018eel, Adhikary:2018ise}.
}

In extensions of two Higgs doublet models with vectorlike quarks, the decays of vectorlike quarks may easily be dominated by cascade decays through charged or neutral Higgs bosons leading to signatures with 6 top or bottom quarks~\cite{Dermisek:2019vkc}, see figure~\ref{fig:diagrams}. The heavy Higgs bosons are effectively pair produced with QCD size cross sections and lead to final states with very small irreducible SM background (the dominant background happens to originate from QCD multi-jet final states). Thus searching for them in top- and bottom-rich events presents a unique opportunity for the LHC. Depending on model parameters and especially the hierarchies in the masses of new quarks and Higgs bosons, many decay chains are possible, and several of them can be simultaneously sizable. Optimal searches for individual possibilities could be designed but the large number of them makes this approach impractical. However, since top quark decays also contain bottom quarks, a common signature for many possible decay chains is 6 bottom quarks in the final state.

We present a search strategy focusing on the $6b$ final state and find the mass ranges of vectorlike quarks and Higgs bosons that can be explored at the LHC. Although the strategy is tailored for the $b_4 \to bH \to bbb$ process we find it to be very effective for all the processes in figure~\ref{fig:diagrams}.  Among other results, the sensitivity to the charged and neutral Higgs bosons, extending to about 2 TeV, stands out when compared to models without vectorlike matter. 

Complementary signatures of cascade decays of heavy neutral Higgs bosons through vectorlike and SM quarks, relevant when vectorlike quarks are lighter than the new Higgs bosons, were recently studied in ref.~\cite{Dermisek:2019heo}. Besides the opposite required hierarchy in masses, the main difference from the decay modes discussed here is that the flavor violating decays $H\to t_4 t$ and $b_4 b$, resulting from Yukawa couplings that mix vectorlike and SM quarks, have to compete with the usual decay modes of heavy Higgs bosons, $H \to \bar t t$ and $\bar b b$. Therefore the branching ratios of heavy Higgses and thus the sensitivity to these channels highly depends on the size and the structure of Yukawa couplings of vectorlike quarks. On the other hand, the decays of the lightest vectorlike quark are necessarily flavor violating. Thus, if the decay channels through heavy Higgs bosons are kinematically open, these can dominate even for very small values of all Yukawa couplings. Large branching ratios for $t_4 \to Ht,\; H^\pm b$ and $b_4 \to Hb,\; H^\pm t$, even close to 100\%, are abundant in random scans of these couplings~\cite{Dermisek:2019vkc}. 

Similar signatures of heavy Higgses and vectorlike leptons were studied in refs.~\cite{Dermisek:2015oja, Dermisek:2015vra, Dermisek:2015hue, Dermisek:2016via, CidVidal:2018eel}.\footnote{Many other signatures of vectorlike quarks and leptons or heavy Higgses are possible. See, for instance, refs.~\cite{Banerjee:2016wls, Das:2018gcr, Coleppa:2019cul}.} A simple embedding into grand unified theories suggests models with both vectorlike quarks and leptons.  The supersymmetric extension with a complete vectorlike family also provides a possibility to understand the values of all large couplings in the SM  from the IR fixed point structure of the renormalization group equations~\cite{Dermisek:2018ujw, Dermisek:2017ihj, Dermisek:2018hxq}. This possibility, that requires both vectorlike quarks and charged and neutral Higgs bosons near the TeV scale, possibly within the reach of the LHC, is the main motivation for the study presented in this paper.

The same or very similar signatures can also be found in composite Higgs models or models with various top partners. For recent analyses, see for example refs.~\cite{Mrazek:2011iu, Xie:2019gya, Benbrik:2019zdp, Cacciapaglia:2019zmj}. 
A detailed study of the $6t$ final state, motivated by composite Higgs models,  that focuses on the presence of up to three high transverse momentum leptons has been presented in ref.~\cite{Han:2018hcu}. This signature is identical  to one of the signatures of vectorlike quarks  in 2HDM, see the bottom left process in figure~\ref{fig:diagrams}. In addition, these signatures can arise in models with $Z^\prime$ or $W^\prime$ with the neural (charged) Higgs boson in figure~\ref{fig:diagrams} replaced by $Z^\prime$ ($W^\prime$), e.g. the $Z^\prime$ with couplings to $b_4$ and $b$ was suggested in connection with the Z-pole anomalies~\cite{Dermisek:2011xu, Dermisek:2012qx} or tensions in rare $B$ decays~\cite{Kawamura:2019rth, Kawamura:2019hxp}. However, the rates for the final states in other models might not reach the rates that are possible in the 2HDM. Finally, we mention that the $6b$ final state has been studied in ref.~\cite{Papaefstathiou:2019ofh} in the context of SM triple Higgs production at future colliders which has kinematics very different from the signatures considered in this work. 

This paper is organized as follows. In Sec.~\ref{sec:decays} we discuss the decay modes in figure~\ref{fig:diagrams} focusing on the branching ratios to various final states expected in the model. In Sec.~\ref{sec:signatures} we suggest search strategies for final states with six bottom quarks and estimate the reach of the LHC for heavy Higgs bosons and vectorlike quarks. We summarize the main results in Sec.~\ref{sec:conclusions}. Supplementary details of the analysis are provided in appendices.

\section{Decays of vectorlike quarks through heavy Higgs bosons}
\label{sec:decays}

Decays of vectorlike quarks through heavy Higgs bosons are especially well motivated, since in 2HDMs they can dominate for a generic set of even very small Yukawa couplings that mix vectorlike and SM quarks. This has been studied in detail  in ref.~\cite{Dermisek:2019vkc}, where a complete extension of the type-II 2HDM by vectorlike quarks with the same quantum numbers as SM quarks (and corresponding conjugate states) was considered. A description of the model, all the relevant formulas for couplings of Higgs and gauge bosons together with a detailed discussion of cascade decays of vectorlike quarks (assuming they dominantly mix with the third generation of SM quarks)  can be found there. Here we summarize the findings at the level sufficient for the motivation of signatures and basic understanding of the results. In addition, we provide new results for scenarios where decay chains through all heavy Higgs bosons are simultaneously kinematically open.

\begin{figure}[t]
\begin{minipage}{6in}
\centering
	\raisebox{-0.5\height}{ \includegraphics[scale=0.35]{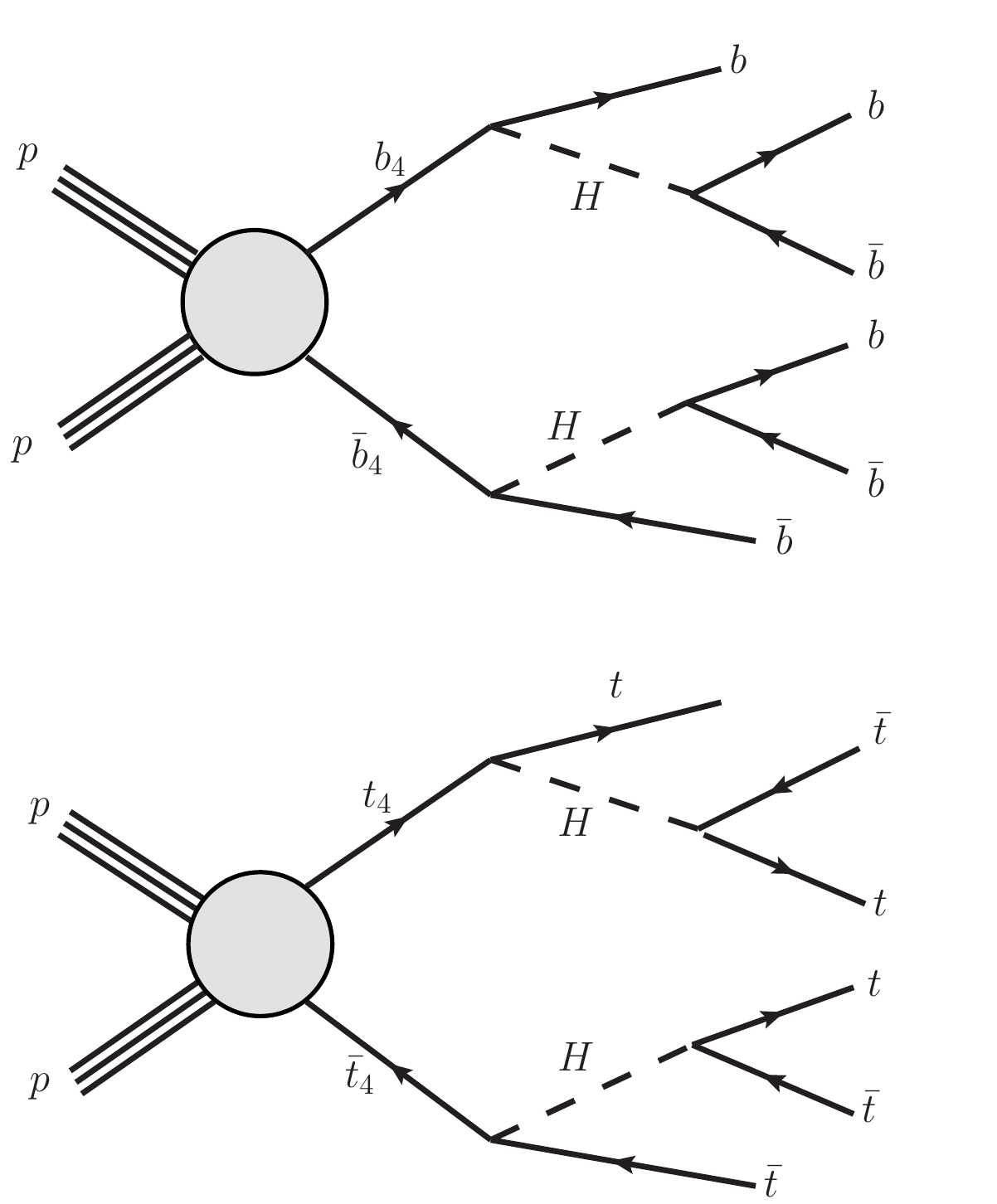} }
  \hspace*{0.1in
  	\raisebox{-0.5\height}{ \includegraphics[scale=0.35]{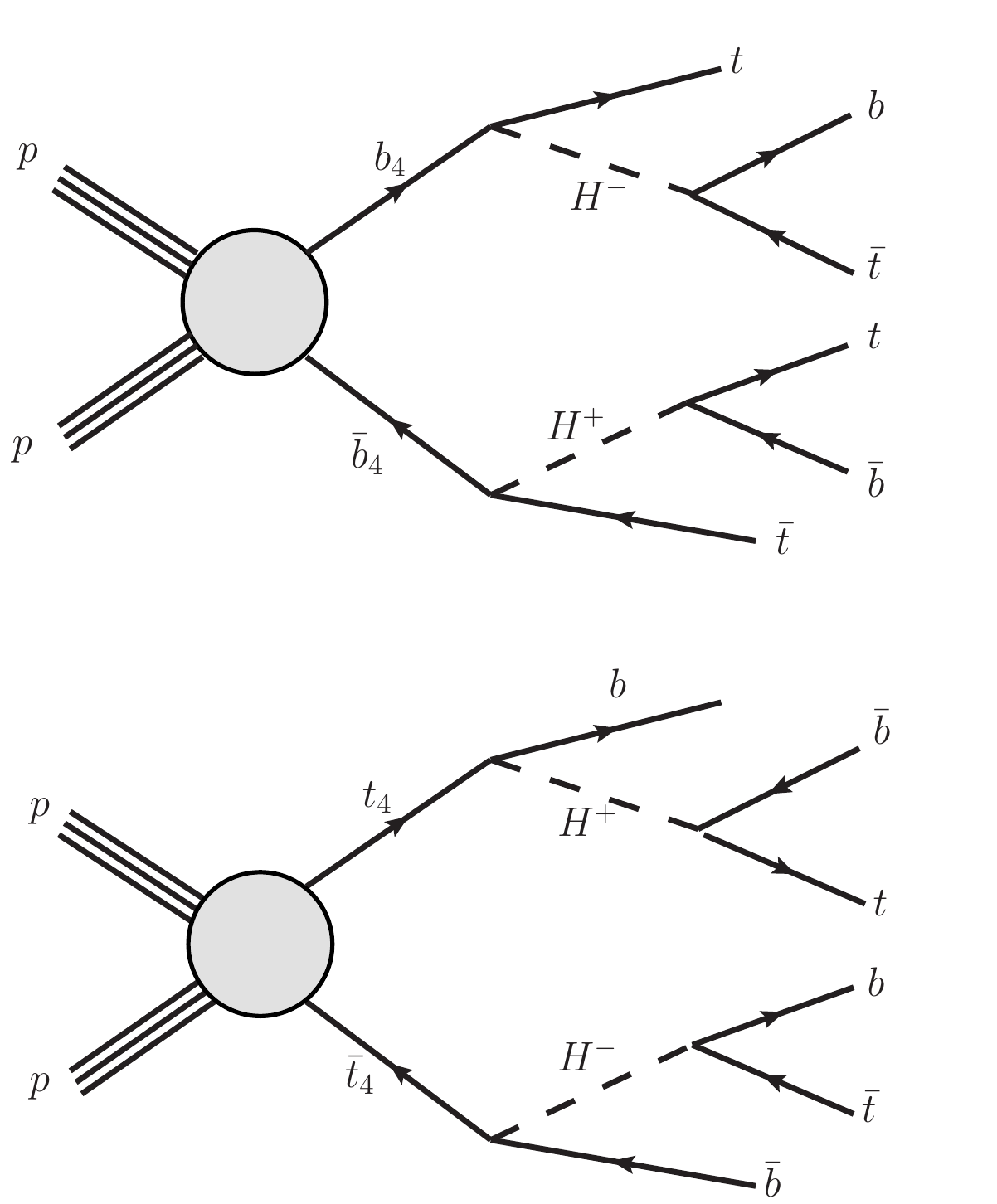} }
    \hspace*{0.1in}
	\raisebox{-0.5\height}{ \includegraphics[scale=0.35]{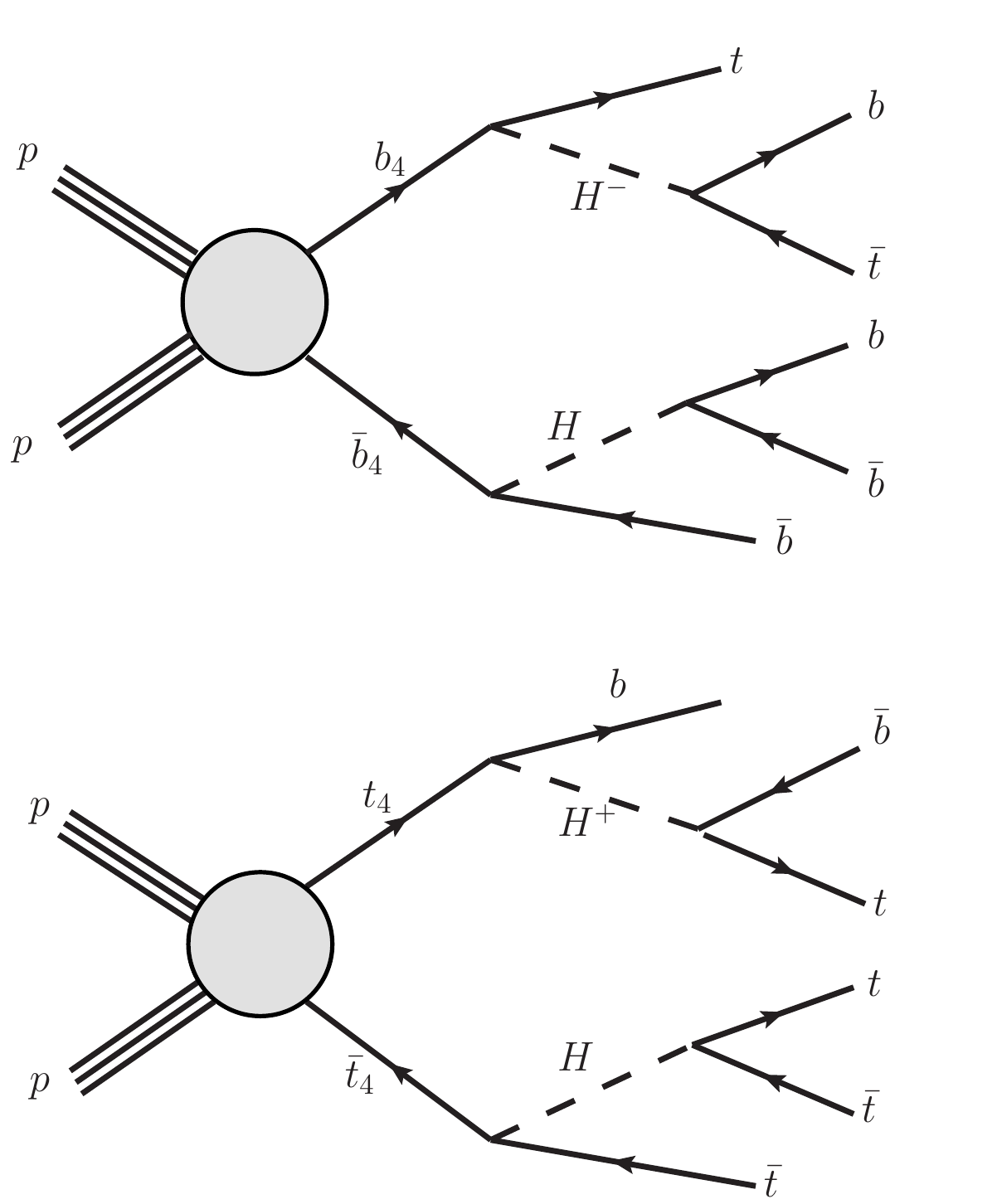}}
\caption{Representative cascade decays of pair produced vectorlike top and bottom quarks through neutral Higgs bosons (left), charged Higgs bosons (middle), and mixed cases of one decaying through a neutral and the other through a charged Higgs boson (right).}
\label{fig:diagrams}
}
\end{minipage}
\end{figure}

In figure~\ref{fig:diagrams} we show representative cascade decays from pair production of a new bottom-like quark, $b_4$, and top-like quark, $t_4$, through charged and neutral Higgs bosons. These include decays of vectorlike quarks through two neutral Higgs bosons, two charged Higgs bosons, and cases when one branch of the decay proceeds through a neutral Higgs and the other through a charged Higgs (only one possibility for each quark flavor is shown).  There are also processes where $H$ in any branch is replaced with $A$. In addition, there are processes with any other allowed decay modes of the individual Higgs bosons. However, the decay chains shown are those that can have combined branching ratio close to 100\% in the  type-II 2HDM. For example, we do not show and will not consider cases such as $t_{4}\rightarrow H t \rightarrow tbb$ or $b_{4}\rightarrow H b \rightarrow ttb$. In these decay chains, the branching ratios of vectorlike quarks and those of the neutral Higgs boson are large in different regions of $\tan\beta$~\cite{Dermisek:2019vkc, Dermisek:2019heo} and the combined branching ratio for the whole chain is not  large for any $\tan\beta$. Thus these topologies are suppressed compared to the ones shown. Note however that in other models such decay chains might be significant.

In order to understand why the decays of vectorlike quarks through heavy Higgs bosons can easily dominate, it is crucial to note that the decays of the lightest new quark are necessarily flavor violating. They can decay into a SM quark and $W$, $Z$, $h$, $H$, $A$, or $H^\pm$. The relevant couplings originate from Yukawa couplings that mix vectorlike (SU(2) doublet or singlet) quarks and SM quarks. The leading dependence of the $t_4$ and $b_4$ partial decay widths on the Yukawa couplings and $\tan \beta$ is summarized in Table 2 of ref.~\cite{Dermisek:2019vkc} (the leading dependences of partial decay widths to $A$ and $H$ are identical). The important observation is that partial decay widths to heavy Higgs bosons are controlled by either a different Yukawa coupling than the partial decay widths to $W$, $Z$, $h$, or they have opposite $\tan\beta$ dependence. Thus, if the decay channels through heavy Higgs bosons are kinematically open, these can dominate even for very small values of all Yukawa couplings. Large branching ratios for $t_4 \to Ht,\; H^\pm b$ and $b_4 \to Hb,\; H^\pm t$, even close to 100\%, are abundant in random scans of these couplings~\cite{Dermisek:2019vkc}.

\begin{figure}[t]
\begin{minipage}{6in}
\centering
	\raisebox{-0.5\height}{ \includegraphics[scale=0.6]{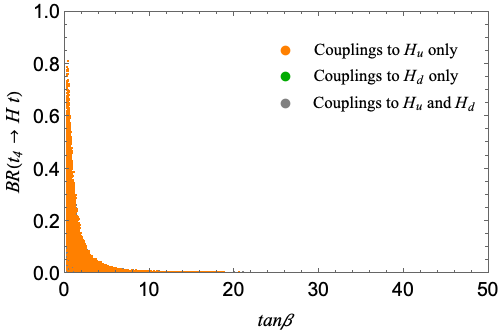}
	 \includegraphics[scale=0.6]{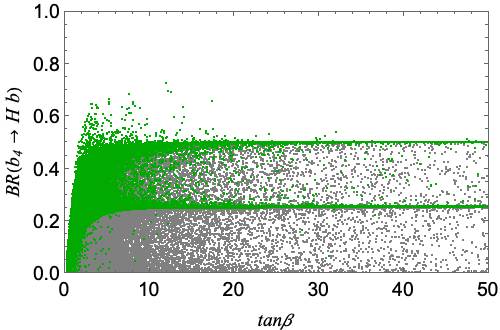} }\\
	\raisebox{-0.5\height}{ \includegraphics[scale=0.6]{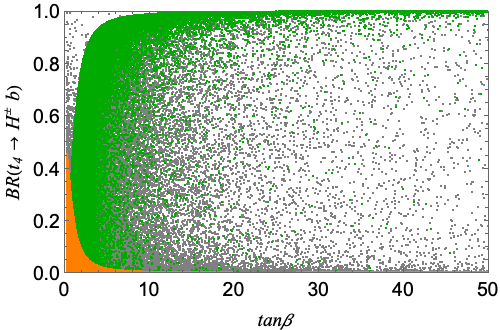}
	 \includegraphics[scale=0.6]{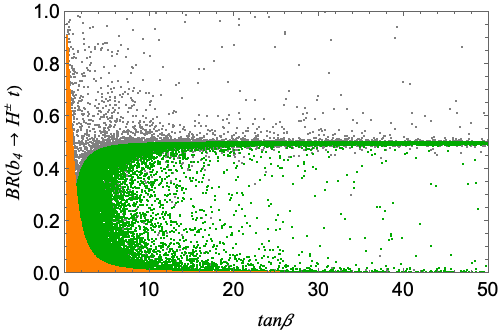}}	
\caption{Allowed branching ratios of $t_4$ and $b_4$ into heavy charged and neutral Higgses vs. $\tan \beta$. In all panels, orange (green) points correspond to a parameter scan assuming couplings to $H_u$ ($H_d$) only; gray points correspond to a parameter scan when all couplings are allowed. In all figures, decay modes through any heavy Higgs boson are allowed and $m_H = m_A = m_{H^\pm} = 1$ TeV is assumed. Other details of the parameter space scan are the same as in ref.~\cite{Dermisek:2019vkc}.}
\end{minipage}
\label{fig:branching_ratios}
\end{figure}

\begin{figure}[t]
\begin{minipage}{6in}
\centering
	\raisebox{-0.5\height}{ \includegraphics[scale=0.5]{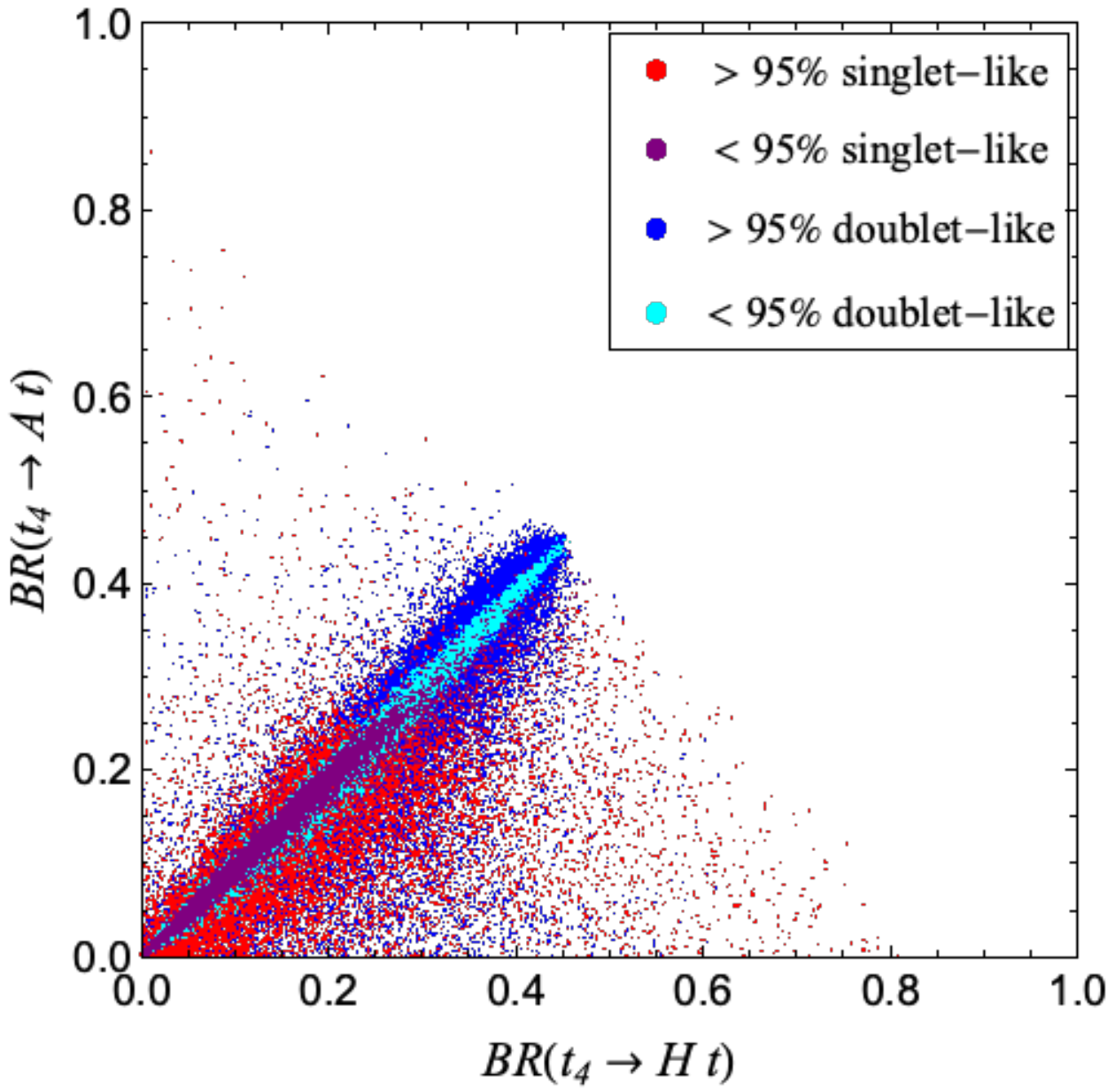}
	 \includegraphics[scale=0.5]{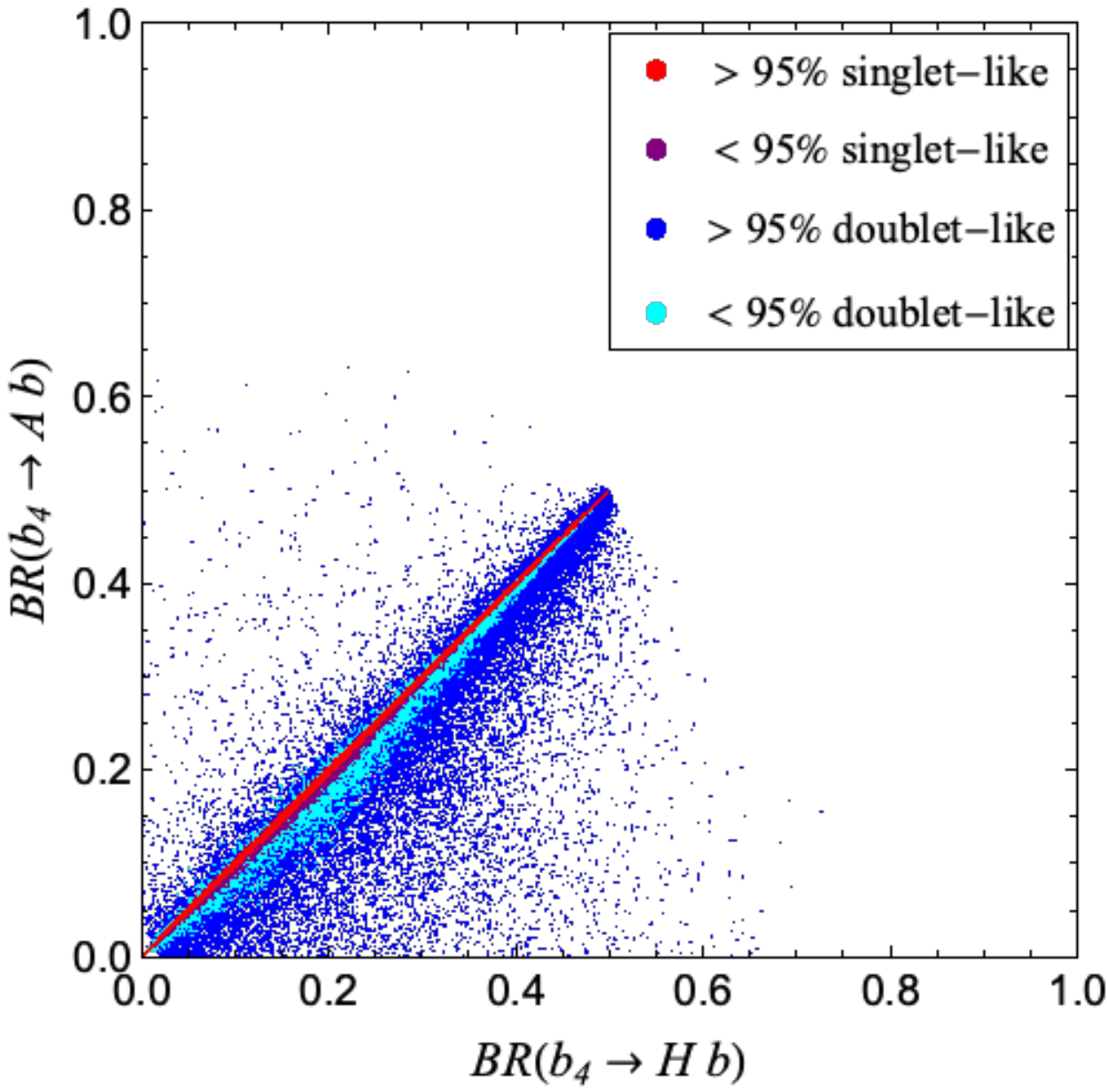} }\\
	\raisebox{-0.5\height}{ \includegraphics[scale=0.5]{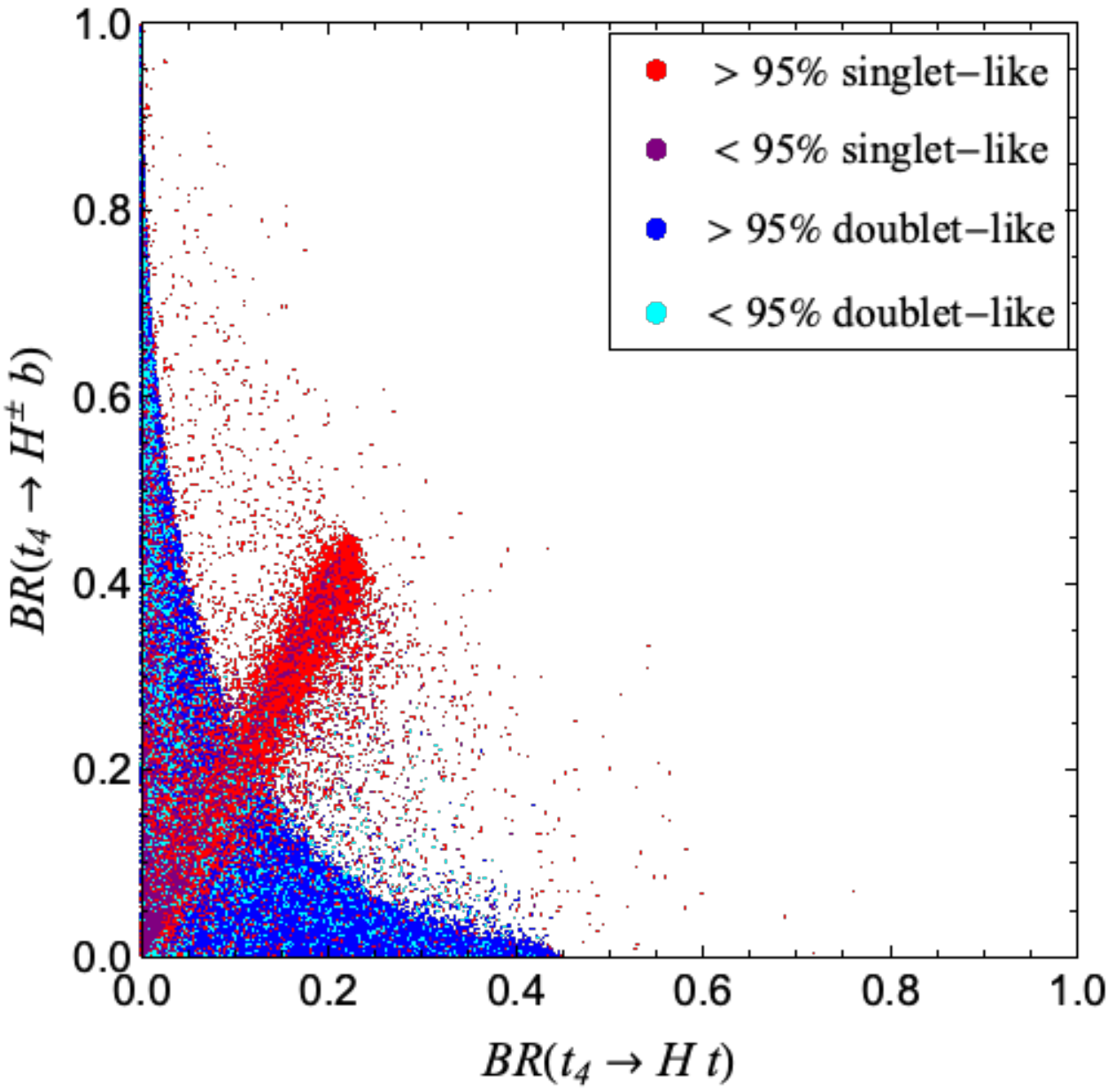}
	 \includegraphics[scale=0.5]{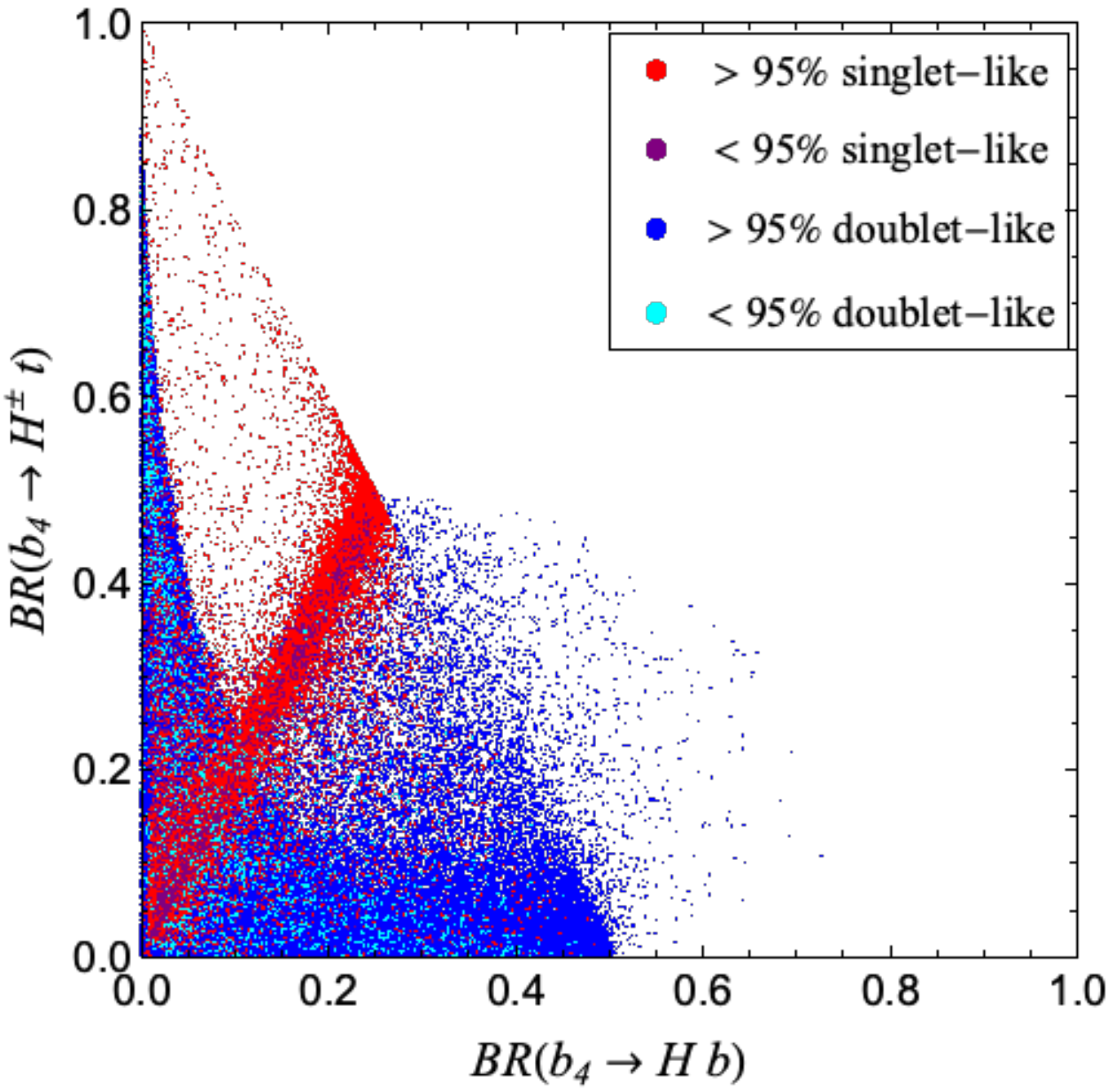}}	
\caption{The allowed branching ratios of $t_4$ and $b_4$ into heavy charged and neutral Higgses in the general scenario with all couplings allowed. The new quark is 95\% or more singlet-like (red),  50\%-95\% singlet-like (purple), 50\%-95\% doublet-like (cyan) or 95\% or more doublet-like (blue). In all figures, decay modes through any heavy Higgs boson are allowed and $m_H = m_A = m_{H^\pm} = 1$ TeV is assumed. Other details of the parameter space scan are the same as in ref.~\cite{Dermisek:2019vkc}.
}
\label{fig:branching_ratios_doublet_singlet}
\end{minipage}
\end{figure}

The analysis in ref.~\cite{Dermisek:2019vkc} assumed that decays to only one of the heavy Higgs bosons were kinematically open. Since some of the topologies in figure~\ref{fig:diagrams} involve decays to two different Higgs bosons, we extend the analysis to the case when decays through all three heavy Higgs bosons are kinematically open. 
Such possibilities are expected to occur in supersymmetric extensions where heavy Higgs bosons are typically almost degenerate. 
In figure~\ref{fig:branching_ratios} we plot the allowed branching ratios of $t_4$ and $b_4$ into heavy charged and neutral Higgses vs. $\tan \beta$ resulting from a random scan over the parameter space described in ref.~\cite{Dermisek:2019vkc} with the additional assumption that $m_H = m_A = m_{H^\pm} = 1$ TeV (the actual mass is not important as long as the decays remain kinematically open). In addition, in figure~\ref{fig:branching_ratios_doublet_singlet} we plot the generated scenarios in planes of various branching ratios. For a vast majority of the points the branching ratio to $A$ is almost identical to the branching ratio to $H$, as is indicated in the top plots in figure~\ref{fig:branching_ratios_doublet_singlet} and thus we do not plot its $\tan \beta$ dependence.\footnote{The differences between couplings of $H$ and $A$ originate from non-holomorphic couplings $\bar \lambda$ and $\bar \kappa$, see eqs.~(A.43), (A.48), (A.49)  and (A.55), (A.57), (A.58) in ref.~\cite{Dermisek:2019vkc}. Effects of these couplings are typically negligible, except for special cases when the leading coupling controlling the branching ratio is not present or is very small compared to others (as indicated by the density of points). In the minimal supersymmetric model extended by vectorlike quarks such couplings are absent and thus the branching ratios would be identical.} The color coding of the points allows for easy understanding of numerical results and it is the same as in ref.~\cite{Dermisek:2019vkc}.

 In all cases, the behavior of the branching ratios and their dependence on $\tan\beta$ is straightforward to understand from their simple relationships to the couplings summarized in Table 2 of ref.~\cite{Dermisek:2019vkc}. We find that, when decay channels through all three heavy Higgses are kinematically open and dominate, the typical ratios of partial widths of $t_4$ to $H$, $A$ and $H^\pm$ are 1/2\,:\,1/2\,:\,0 (doublet-like $t_4$) or $1/4\,:\,1/4\,:\,1/2$ (singlet-like $t_4$) at very small $\tan \beta$, and $0\,:\,0\,:\,1$ (doublet-like $t_4$) for medium to large $\tan \beta$.  The ratios of partial widths of $b_4$ to $H$, $A$ and $H^\pm$ are $0\,:\,0\,:\,1$ (doublet-like $b_4$) at very small $\tan \beta$,  and $1/4\,:\,1/4\,:\,1/2$ (singlet-like $b_4$) or $1/2\,:\,1/2\,:\,0$ (doublet-like $b_4$) for medium to large $\tan \beta$.  

In summary, we find that the branching ratios of $t_4 \to H^\pm b \to tbb$ and $b_4 \to H (A) b \to bbb$ can easily be close to 100\% for any medium to large $\tan\beta$ even if all the decay modes are kinematically open. Thus these decay modes are especially well motivated. In the following section we focus on the $6b$ final state and estimate the sensitivity of the LHC to $b_4$ and $H(A)$. We also use the same analysis to find sensitivities for other decays including $t_4 \to H^\pm b \to tbb$.

\section{Search strategies and reach at the LHC}
\label{sec:signatures}
The final states shown in figure~\ref{fig:diagrams} are $6b$ ($b_4 \to H$ cascade),  $6t$ ($t_4\to H$ cascade), $4b2t$ and $4t2b$ (both requiring the presence of $H^\pm$). In this section we develop an analysis strategy, based on the presence of four or five $b$-jets with high transverse momentum and large total transverse energy, which applies to all these final states. Obviously this strategy is expected to be the most effective for the $6b$ final state, with slightly weaker sensitivity as the number of top-jets increases.

\begin{figure}[ht]
\centering
	\includegraphics[scale=0.75]{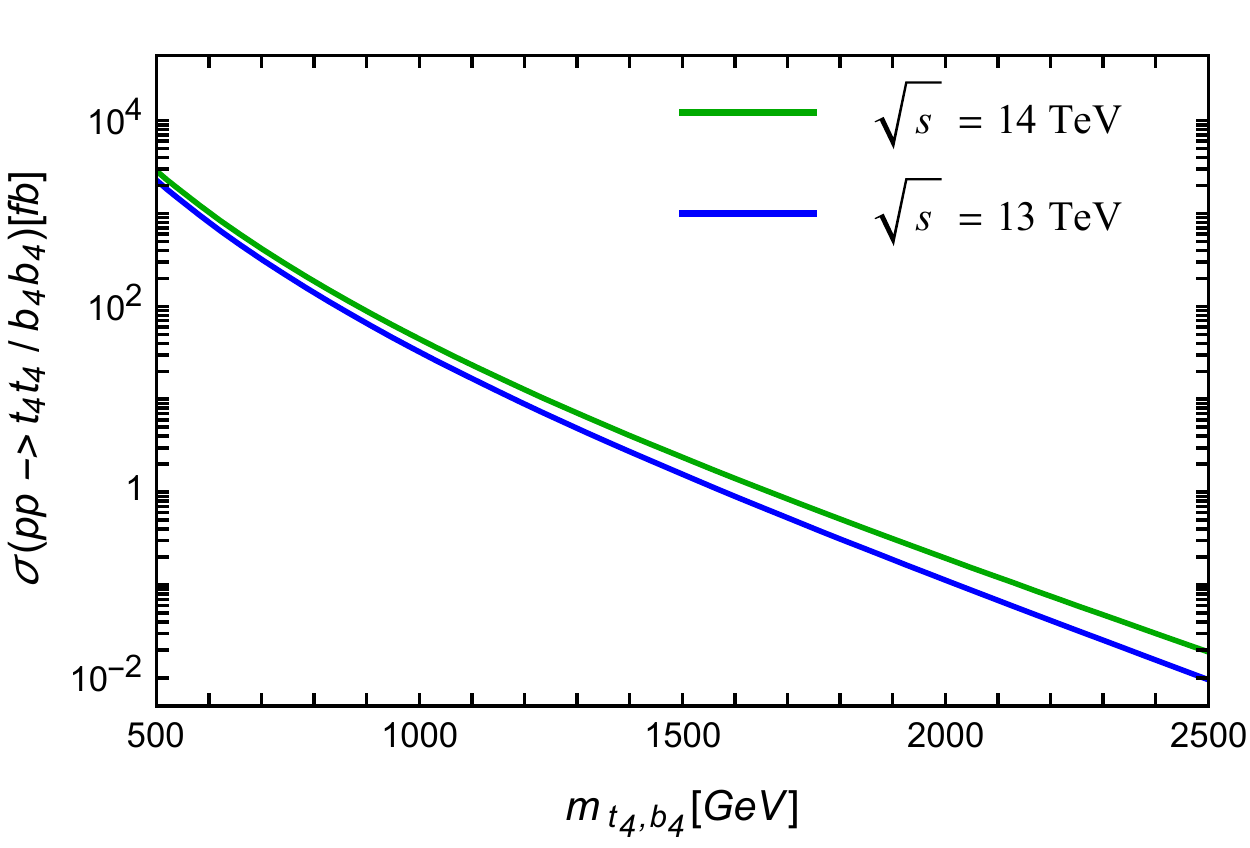}
\caption{Production cross section for pair-produced vectorlike top or bottom quarks at the LHC. In green (blue) we show the LO cross section for 14 TeV (13 TeV) center of mass energy.}
\label{fig:prod_xsection}
\end{figure}

In figure~\ref{fig:prod_xsection} we show the LO production cross section of pair produced vectorlike quarks at the LHC as a function of their masses $m_{t_{4},b_{4}}$. We show curves for 14 TeV  and 13 TeV center of mass energies in green and blue, respectively. 

To generate each signal, we implement the model discussed in ref.~\cite{Dermisek:2019vkc} into {\tt FeynRules}~\cite{Degrande:2011ua} to produce a {\tt UFO} file. Parton level events for signal and background are then generated with {\tt MadGraph5}~\cite{Alwall:2014hca} and subsequently showered and hadronized with {\tt Pythia8}~\cite{Sjostrand:2006za, Sjostrand:2014zea}. Finally we used {\tt Delphes}~\cite{deFavereau:2013fsa} (with standard settings) to simulate detector effects. The last step is imperative to our analyses as a crucial aspect of the search strategy relies on the estimation of the efficiency to observe multiple $b$-jets with high transverse momentum.

A useful kinematical quantity is the total transverse energy of reconstructed $b$-tagged jets:
\begin{align}
H_{Tb} \equiv \sum_{j\ni b} |p_T(j)|\; ,
\end{align}
which for vectorlike quark masses above 1 TeV can easily exceed 2-3 TeV. Cutting on $H_{Tb}$ allows us to strongly reduce multi-jet QCD backgrounds while, at the same time, preserving sizable signal efficiency.

The dominant obstacle to an all-hadronic analysis is the estimation of the QCD background. In particular, besides irreducible backgrounds with at least four or five $b$ quarks at the parton level (e.g. $p p \to 4b$), there are multi-jet final states in which regular jets are mistagged as $b$ (e.g. $p p \to 2b2j$). The mistag rate receives a small contribution ($1-2\%$) from jets which do not contain $b$-hadrons after hadronization but are nevertheless tagged by the algorithm, and a much larger one (we found about 7\% from a study based on simulations with {\tt Pythia8}) from jets in which a gluon splits into a $b\bar b$ pair. The parton level cross sections for processes in which a pair of $b$ quarks is replaced with two jets receive an enormous combinatoric enhancement and, after detector simulation, dominate the background to a multi $b$-jets signal. The problem of separating genuine $b$-jets (which we refer to as ``1b'') from jets in which a gluon splits into $b\bar b$ (which we refer as ``2b'') has been already investigated in the literature~\cite{Behr:2015oqq, Goncalves:2015prv, Ridolfi:2019bch, Sirunyan:2020hwv}. For instance, the effect we are describing has been discussed in the theoretical investigation presented in ref.~\cite{Behr:2015oqq} in the context of background for $pp \to hh \to 4b$ where an effective mistag rate of about 6\% is found. On the experimental side, a recent search for a charged Higgs  by CMS~\cite{Sirunyan:2020hwv} found that the background from misidentified $b$-jets dominates over the genuine $b$-jets one. More interestingly, in ref.~\cite{Goncalves:2015prv} the authors presented a tagging algorithm to discriminate between ``1b'' and ``2b'' jets and showed that, combining girth, charged track multiplicity and momentum fraction of the leading $b$-hadron, it is possible to achieve a sizable tagging efficiency for jets containing one $b$-hadron $(\epsilon_{1b})$ while having at the same time a large rejection of jets containing two $b$-hadrons $(\bar{\epsilon}_{2b})$. In our analysis we adopt their tagging strategy but, instead of implementing directly their tagging algorithm, we simply adopt one average working point and re-weight signal and background events accordingly (details of this procedure are discussed in appendix~\ref{app:rescaling}). Note that, according to ref.~\cite{Goncalves:2015prv}, the tagging algorithm becomes more efficient with increasing $p_T$ of the jets. Since our cuts enforce the presence of very high $p_T$ $b$-jets, we expect that our implementation is conservative.

In the next two subsections we discuss the reach of two possible search strategies based on tagging at least four and five $b$-jets.  Both analysis strategies that we propose require $b$-tagged jets with $\Delta R > 0.5$, $|\eta| < 3$, $p_T > p_T^{\rm cut}$ and $H_{Tb} > H_{Tb}^{\rm cut}$. The actual values of the $p_T$ and $H_{Tb}$ cuts are optimized for a given choice of vectorlike quark and heavy Higgs mass. Typical values are $p_T^{\rm cut} \in [100,300]$ GeV and $H_{Tb}^{\rm cut}\in [1,3]$ TeV. We do not require the four (five) $b$-tagged jets to be the four (five) highest-$p_T$ jets in the event; in fact, the $b$-tagging efficiency decreases at large $p_T$. We consider the four (five) highest $p_T$ $b$-jets in each event in order to take advantage of the combinatorics (all our signals have at least six $b$-jets) and thus achieve a larger signal efficiency. For the 4b and 5b analyses we adopt $(\epsilon_{1b} ,\bar\epsilon_{2b}) = (0.8,1/3)$ and $(\epsilon_{1b} ,\bar\epsilon_{2b}) = (0.9,1/2)$, respectively. In the 5b analysis we additionally require cuts on the invariant mass of two and three jets.

\begin{figure}[ht]
\centering
	\includegraphics[width=0.48\linewidth]{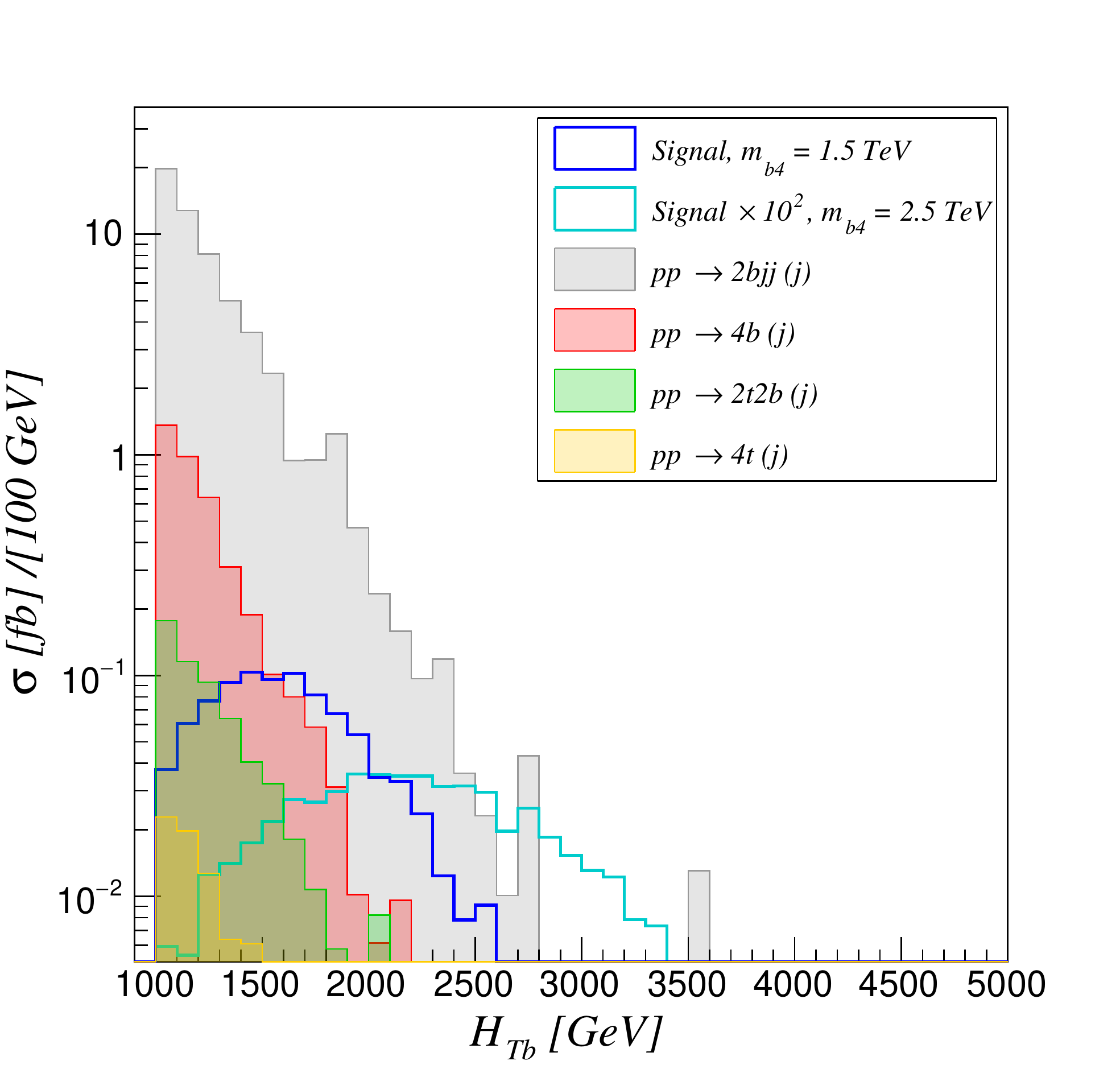} 
	\includegraphics[width=0.48\linewidth]{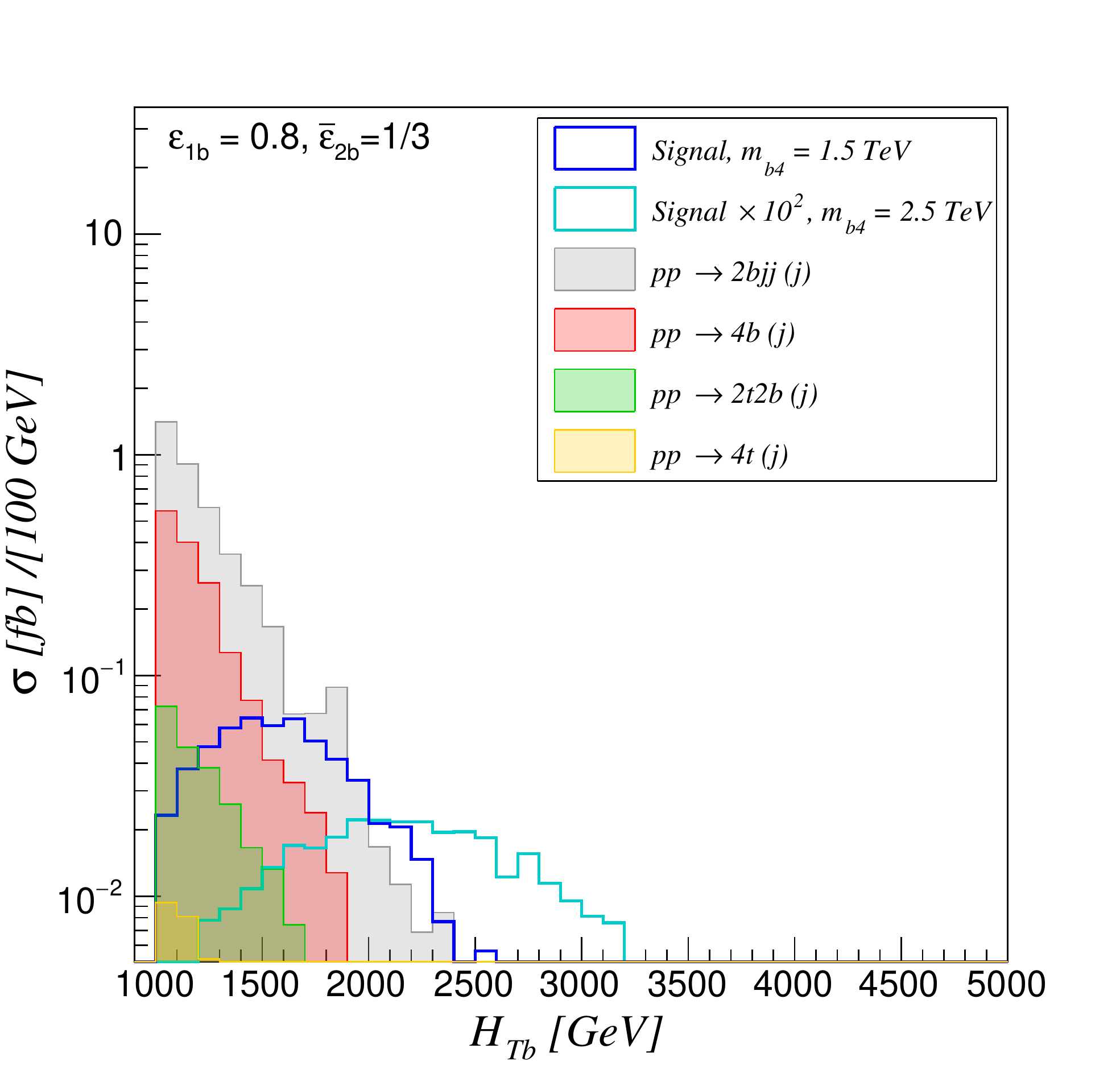} 
\caption{{\bf Left:} $H_{Tb}$ distribution for the $6b$ signal and backgrounds in the 4b analysis. We show the signal distribution for $m_{b_4}=1.5$ and 2.5 TeV in blue and cyan respectively. The gray, red, green, and yellow shaded regions show the $H_{Tb}$ distributions of the $2bjj(j)$ $4b(j)$, $2t2b (j)$, and $4t (j)$ backgrounds respectively. {\bf Right:} $H_{Tb}$ distribution for the $6b$ signal and backgrounds after incorporating the 1b/2b-tagging strategy assuming $(\epsilon_{1b} ,\bar\epsilon_{2b}) = (0.8,1/3)$ efficiencies.}
\label{fig:4b_HTb}
\end{figure}

Finally, let us comment on the reason for adopting $H_{Tb}$ rather than the total hadronic transverse energy $H_T$. Large $H_T$ multi-jet backgrounds are very difficult to simulate if the minimum jet transverse momentum ($p_T^{\rm cut}$) is much smaller than $H_T^{\rm cut}$. In fact, configurations with multiple relatively low-$p_T$ and well separated\footnote{Splitting a jet leads to a much larger $H_T$ only if the two resulting jets have large $\Delta R$.  Shower generated jets do not capture these effects since QCD radiation is calculated in the collinear limit.} jets are enhanced with respect to those with fewer higher-$p_T$ jets. As a result one needs to include final states where the number of well-separated jets is roughly up to $H_T^{\rm cut}/p_T^{\rm cut}$. Small scale computer simulations allow for up to 6 jets in the final states implying that QCD background with $H_T^{\rm cut}  \gtrsim 1$ TeV and $p_T^{\rm cut} \gtrsim 100$ GeV is dangerously sensitive to higher order corrections. A possible solution to this problem is to refrain from a calculation and use signal-depleted regions to measure the multi-jet background. Another strategy, which we follow here in order to give an idea of the sensitivity that this kind of analyses can yield, is to replace $H_T$ with $H_{Tb}$. Finally, an alternative method is to perform a multivariate analysis which uses the transverse momentum of all identified $b$-jets as input.

\subsection{4b analysis}
\label{sec:4b}
\begin{figure}[t]
\centering
	\includegraphics[scale=0.34]{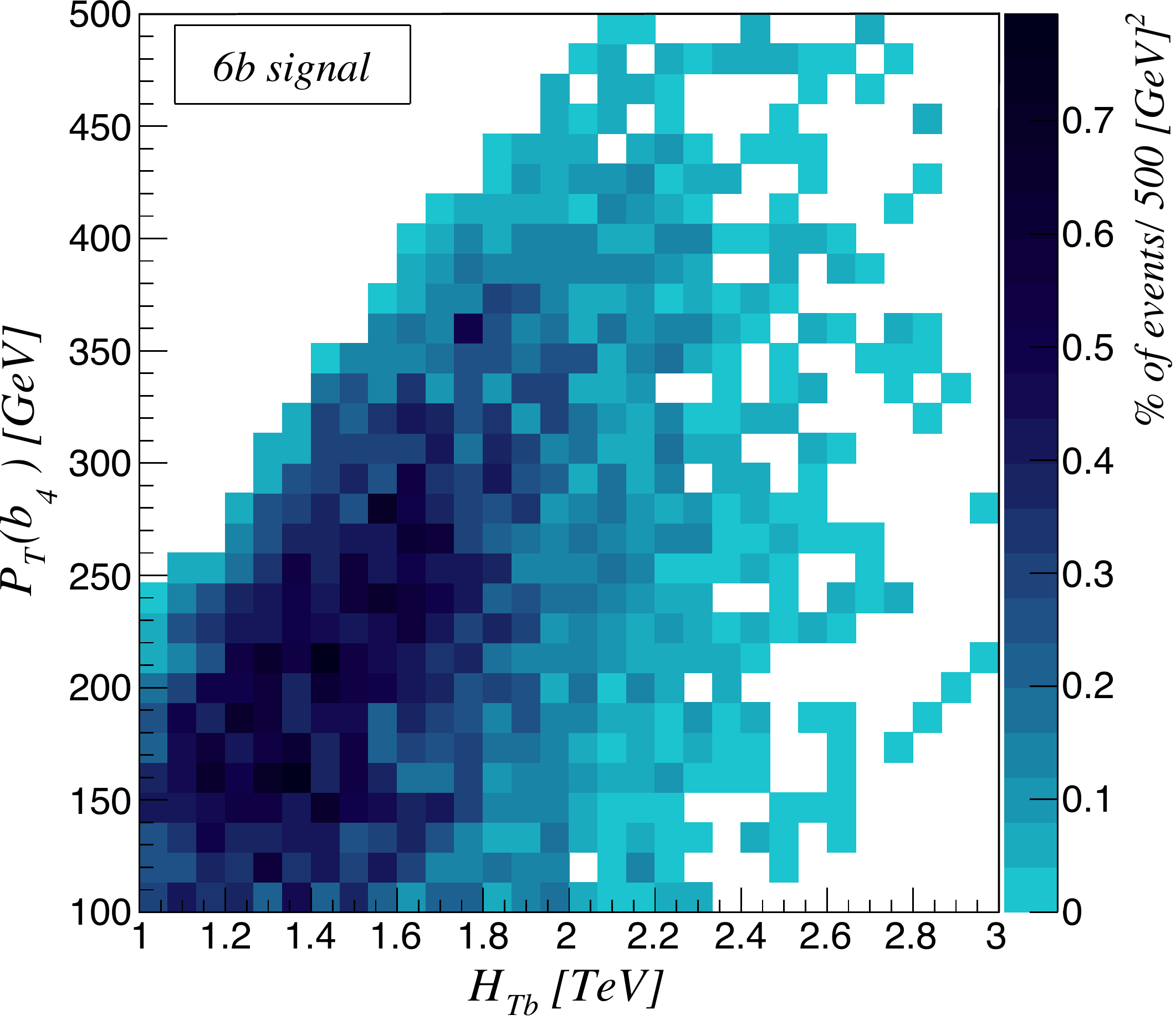} 
		\includegraphics[scale=0.34]{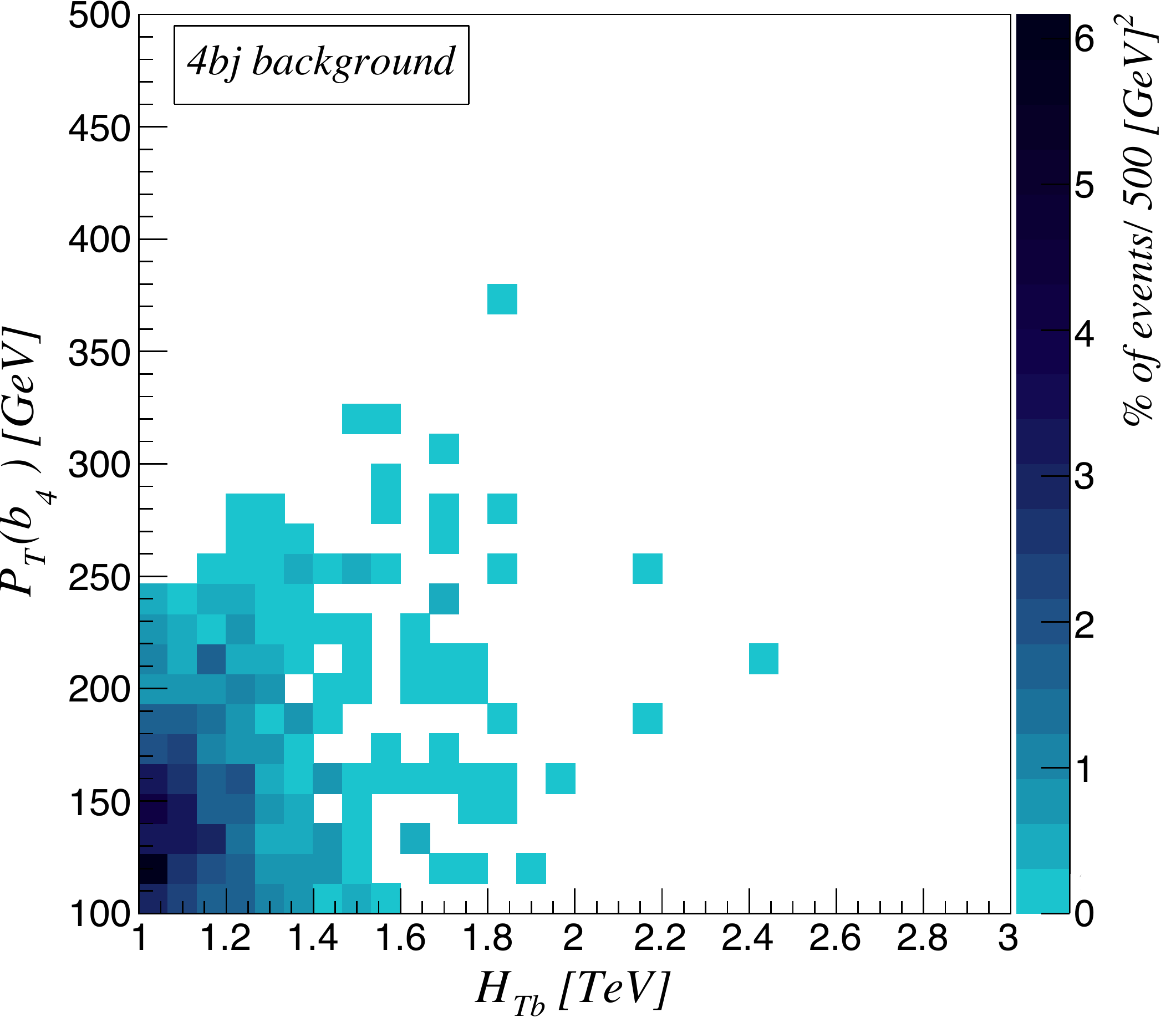} 
\caption{{\bf Left}: Distribution of events for the $6b$ signal in the $[H_{Tb},P_{T}(b_{4})]$ plane for $m_{b4} = 1.5$ TeV. {\bf Right:} Distribution of events for the $4bj$ background in the $[H_{Tb},P_{T}(b_{4})]$ plane. In both panels, $P_{T} (b_4)$ is the transverse momentum of the fourth-highest $p_{T}$ identified $b$-jet.}
\label{fig:HTb_PT4}
\end{figure}

The dominant backgrounds to an analysis based on four high-$p_T$ $b$-tagged jets are ($2b2j$, $4b$, $4t$, $2b2 t$) + n-jets ($n = 0,1$). In figure~\ref{fig:4b_HTb} we show the $H_{Tb}$ distribution of backgrounds and $6b$ final state signals for $m_{b_4} =$ 1.5 and 2.5 TeV. The panel on the right shows the impact of adopting the 1b/2b tagging strategy, whose main effect is to suppress the $2bjj(j)$ background by an order of magnitude. In figure~\ref{fig:HTb_PT4} we show the distributions of $H_{Tb}$ versus the transverse momentum of the fourth-highest $p_{T}$ identified $b$-jet, $P_{T} (b_4)$, for the $m_{b_4} = 1.5$ TeV signal and $4b$ + n-jets background. The requirement of large $H_{Tb}$ implies the presence of several high-$p_T$ $b$-tagged jets. The softest identified $b$-jet is usually much harder in our signal (due to the presence of heavy resonances) rather than in the QCD background. Therefore, since in the background large $H_{Tb}$ is achieved mostly from the leading jets, cutting on $P_{T} (b_4)$ offers a good discrimination.

In tables~\ref{tab:4bAnalysis_b4} and \ref{tab:4bAnalysis_t4} we present the result of the $(p_T^{\rm cut}, H_{Tb}^{\rm cut})$ optimization. For a range of $m_{b_4,t_4}$ masses we give the optimal cuts, the fiducial signal cross section assuming 100\% branching ratios and fiducial background cross sections. All cross sections are calculated for $pp$ collisions at 14 TeV. The results in these tables can be easily converted into upper limits onto branching ratios for vectorlike quark decays into neutral and charged Higgses following the statistical procedure detailed in appendix~\ref{app:poisson}. 

\begin{figure}[ht]
\centering
	\includegraphics[scale=0.85]{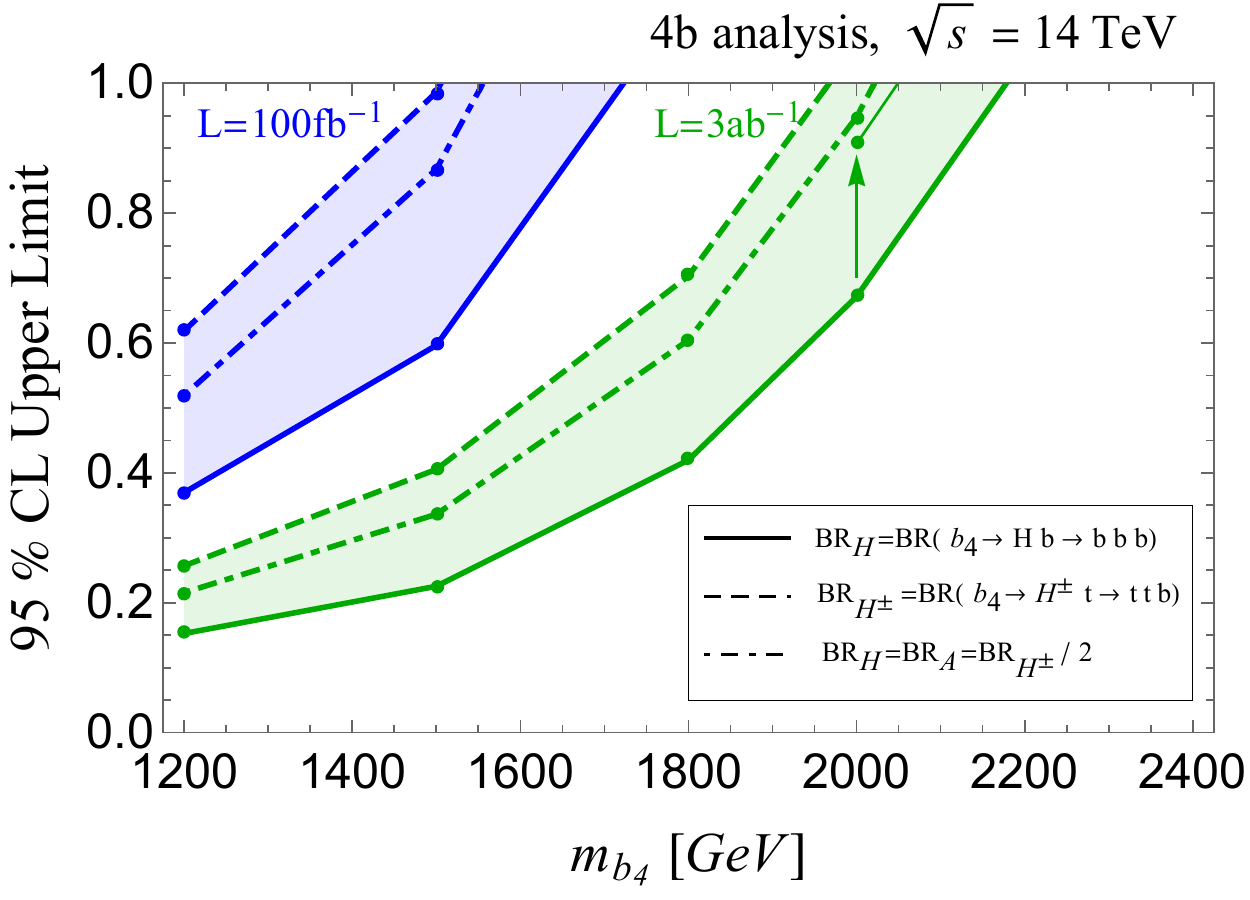}  
\caption{Expected $95\%$ CL upper limits on the $b_4$ branching ratio into neutral or charged Higgses in the 4b analysis. Upper limits for the $b_4\rightarrow H b\rightarrow bbb$ decay mode are shown with the solid lines, while those for the  $b_4\rightarrow H^{\pm} t\rightarrow ttb$ are shown with the dashed lines assuming $m_{H,A,H^{\pm}}=1$ TeV. The dot-dashed lines give the limits when all heavy Higgses decay with typical patterns of branching ratios for a singlet-like $b_{4}$ as discussed in section~\ref{sec:decays} with $m_{H,A,H^{\pm}}=1$ TeV. The thinner line with the arrow indicate the corresponding limit when $m_{H,A,H^{\pm}}=m_{b_{4}} - 200$ GeV.}
\label{fig:4b_b4_bounds}
\end{figure}

Before discussing the sensitivities we obtain, let us comment on the contributions of the $4b$ + {n-jets} background with $n \geq 2$. We have studied the breakdown of the dominant background $4b(j)$ into $4b$ (without additional jets with $p_T(j) > 100$ GeV and $\Delta R > 0.5$) and  $4b+ j$ (with $p_T > 100$ GeV and $\Delta R > 0.5$). The latter receives contributions from $qg$ initiated hard parton scattering which, after imposing the cuts mentioned above, accounts for about 60\% of the total $4b + j$ background. The $4b + j$ cross section stemming from initial parton configurations that are shared with $4b$ is smaller than the $4b$ cross section. This suggests that the cuts we consider allow for a perturbative estimate of the relevant backgrounds; in particular, we expect the $4b$ + {n-jets} background with $n \geq 2$ to be subdominant.

\begin{figure}[ht]
\centering
	\includegraphics[scale=0.85]{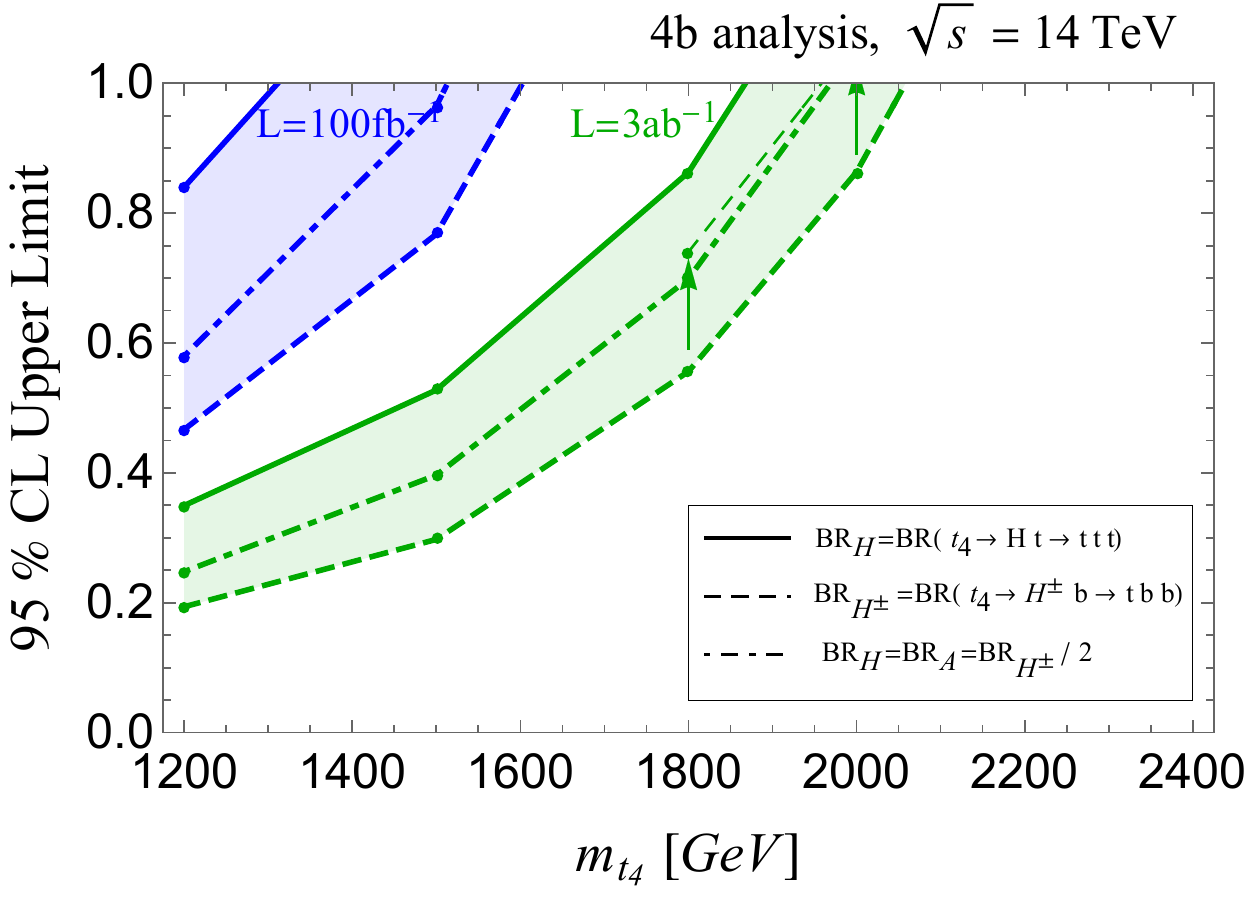} 
\caption{Expected $95\%$ CL upper limits on the $t_4$ branching ratio into neutral or charged Higgses in the 4b analysis. Upper limits for the $t_4\rightarrow H t\rightarrow ttt$ decay mode are shown with the solid lines, while those for the  $t_4\rightarrow H^{\pm} b\rightarrow tbb$ are shown with the dashed lines assuming $m_{H,A,H^{\pm}}=1$ TeV. The dot-dashed lines give the limits when all heavy Higgses decay with typical patterns of branching ratios for a singlet-like $t_{4}$ as discussed in section~\ref{sec:decays} with $m_{H,A,H^{\pm}}=1$ TeV. The thinner line with arrows indicate the corresponding limit when $m_{H,A,H^{\pm}}=m_{t_{4}} - 200$ GeV.}
\label{fig:4b_t4_bounds}
\end{figure}

In figures~\ref{fig:4b_b4_bounds} and \ref{fig:4b_t4_bounds} we present the results that we obtain for $b_4$ and $t_4$ cascade decays. In both figures blue (green) bounds correspond to 100 (3000) fb${}^{-1}$ of integrated luminosity at 14 TeV. For the LHC running at 13 TeV and current luminosity of 139 fb${}^{-1}$ we find that the curves would almost overlap with the bounds presented for 14 TeV 100 fb${}^{-1}$.

For a given $b_4$ ($t_4$) mass we present a limit on the total branching ratio of $b_4$ ($t_4$) into all kinematically accessible heavy Higgses ($H$, $A$ and $H^\pm$). If only one Higgs channel is open the possible final states are $6b$ ($6t$), which is mediated by $H$, and $4t2b$ ($4b2t$), which is mediated by $H^\pm$. The limits we obtain on ${\rm BR} (b_4\to H b)$ (${\rm BR} (t_4\to H t)$) and ${\rm BR} (b_4\to H^- t)$ (${\rm BR} (t_4\to H^+ b)$) assuming $m_{H,H^{\pm}}=1$ TeV are shown as solid and dashed curves, respectively. The thinner lines and arrows indicate how the upper range of the bounds change when $m_{H,H^{\pm}}=m_{t_{4},b_{4}} - 200$ GeV. 

\begin{table}[h]
\begin{center}
\resizebox{!}{1.5in}{%
 \begin{tabular}{|c || c | c | c | c | c |} 
 \hline
 $m_{b_4}$  & 1200 & 1500 & 1800 & 2000 & 2200\\ [0.5ex] 
\hline\hline
$6b$ signal & $5.66\times 10^{-1}$&$1.79\times 10^{-1}$ &$2.99\times 10^{-2}$ &$1.17\times 10^{-2}$ & $4.21\times 10^{-3}$\\
\hline
$\epsilon_{signal}$ & 4.5 \% & 7.5 \% & 5.8 \% & 6.0 \% & 5.6\%\\
\hline
2bjj(j)& $5.46\times 10^{-2}$ & $2.20\times 10^{-2}$ & $5.48\times 10^{-3}$ & $5.48\times 10^{-3}$ & $5.14\times 10^{-3}$ \\
\hline
4b(j)& $5.80\times 10^{-2}$ & $2.89\times 10^{-2}$&$9.63\times 10^{-3}$ &$9.63\times 10^{-3}$ & $5.77\times 10^{-3}$\\
\hline
2t2b(j) & $1.13\times 10^{-2}$& $5.53\times 10^{-3}$ & $1.96\times 10^{-3}$ & $1.96\times 10^{-3}$ & $1.30\times 10^{-3}$\\
\hline
4t(j) & $3.26\times 10^{-4}$& $1.12\times 10^{-4}$ & $5.70\times 10^{-6}$ & $5.70\times 10^{-6}$ & $4.98\times 10^{-6}$\\
\hline
$(H^{cut}_{T_b}, p^{cut}_{T})$& $(1500,200)$ & $(1500,250)$ & $(1750,300)$ & $(1750,300)$& (2500,200)\\
\hline\hline
$4t2b$ signal & $6.24\times 10^{-1}$&$1.41\times 10^{-1}$ &$4.68\times 10^{-2}$ &$2.07\times 10^{-2}$ & $3.91\times 10^{-3}$\\
\hline
$\epsilon_{signal}$ & 4.9 \% & 5.9 \% & 9.1 \% & 11 \% & 5.1\%\\
\hline
2bjj(j)& $7.22\times 10^{-1}$ & $2.19\times 10^{-1}$ & $2.19\times 10^{-1}$ & $2.19\times 10^{-1}$ & $3.56\times 10^{-2}$ \\
\hline
4b(j)& $4.77\times 10^{-1}$ & $1.41\times 10^{-1}$&$1.41\times 10^{-1}$ &$1.41\times 10^{-1}$ & $2.19\times 10^{-2}$\\
\hline
2t2b(j) & $6.75\times 10^{-2}$& $3.44\times 10^{-2}$ & $3.44\times 10^{-2}$& $3.44\times 10^{-2}$ & $6.27\times 10^{-3}$\\
\hline
4t(j) & $6.73\times 10^{-3}$& $2.50\times 10^{-3}$ & $2.50\times 10^{-3}$ &$2.50\times 10^{-3}$ & $2.22\times 10^{-4}$\\
\hline
$(H^{cut}_{T_b}, p^{cut}_{T})$& $(1250,100)$ & $(1500,100)$ & $(1500,100)$ & $(1500,100)$& (1750,200)\\
\hline
\end{tabular}}
\caption{Signals $pp\rightarrow b_{4}b_{4}\rightarrow 6b$,  $pp\rightarrow b_{4}b_{4}\rightarrow 4t2b$ and background fiducial cross sections in units of fb in the 4b analysis. For each $b_{4}$ mass hypothesis, the optimized cuts are listed in the last row for integrated luminosity $L=3\text{ ab}^{-1}$. The signal efficiency of the cuts $(\epsilon)$ is also presented. All results are presented using the 1b/2b-tagging working point $(\epsilon_{1b}, \bar{\epsilon}_{2b})=(0.8,1/3)$.} 
\label{tab:4bAnalysis_b4}
\end{center}
\end{table}

\vspace{0.5cm}

\begin{table}[h]
\begin{center}
\resizebox{!}{1.5in}{%
 \begin{tabular}{|c || c | c | c | c | c |} 
 \hline
 $m_{t_4}$  & 1200 & 1500 & 1800 & 2000 & 2200\\ [0.5ex] 
\hline\hline
$6t$ signal & $3.37\times 10^{-1}$&$1.48\times 10^{-1}$ &$3.11\times 10^{-2}$ &$1.44\times 10^{-2}$ & $6.10\times 10^{-3}$\\
\hline
$\epsilon_{signal}$ & 2.7 \% & 6.2 \% & 6.1 \% & 7.4 \% & 8.0\%\\
\hline
2bjj(j)& $7.22\times 10^{-1}$ & $7.22\times 10^{-1}$ & $2.19\times 10^{-1}$ & $2.19\times 10^{-1}$ & $2.19\times 10^{-1}$ \\
\hline
4b(j)& $4.77\times 10^{-1}$ & $4.77\times 10^{-1}$& $1.41\times 10^{-1}$ & $1.41\times 10^{-1}$ & $1.41\times 10^{-1}$\\
\hline
2t2b(j) & $6.76\times 10^{-2}$& $6.76\times 10^{-2}$ & $3.44\times 10^{-2}$ & $3.44\times 10^{-2}$ & $3.44\times 10^{-2}$\\
\hline
4t(j) & $6.73\times 10^{-3}$ & $6.73\times 10^{-3}$ & $2.50\times 10^{-3}$ & $2.50\times 10^{-3}$ & $2.50\times 10^{-3}$\\
\hline
$(H^{cut}_{T_b}, p^{cut}_{T})$& $(1250,100)$ & $(1250,100)$ & $(1500,100)$ & $(1500,100)$& (1500,100)\\
\hline\hline
$4b2t$ signal & $1.10$ & $1.03\times 10^{-1}$ &$2.94\times 10^{-2}$ &$1.29\times 10^{-2}$ & $2.89\times 10^{-3}$\\
\hline
$\epsilon_{signal}$ & 8.7 \% & 4.3 \% & 5.7 \% & 6.6 \% & 3.8\%\\
\hline
2bjj(j)& $7.22\times 10^{-1}$ & $2.20\times 10^{-2}$ & $2.20\times 10^{-2}$ & $3.56\times 10^{-2}$ & $7.50\times 10^{-3}$ \\
\hline
4b(j)& $4.77\times 10^{-1}$ & $2.89\times 10^{-2}$ & $2.89\times 10^{-2}$ & $2.19\times 10^{-2}$ & $5.76\times 10^{-3}$\\
\hline
2t2b(j) & $6.75\times 10^{-2}$& $5.53\times 10^{-3}$ & $5.53\times 10^{-3}$ & $6.27\times 10^{-3}$ & $2.64\times 10^{-3}$\\
\hline
4t(j) & $6.73\times 10^{-3}$& $1.12\times 10^{-4}$ & $1.12\times 10^{-4}$ &$2.22\times 10^{-4}$ & $1.23\times 10^{-5}$\\
\hline
$(H^{cut}_{T_b}, p^{cut}_{T})$& $(1250,100)$ & $(1500,250)$ & $(1500,250)$ & $(1750,200)$& (2250,200)\\
\hline
\end{tabular}}
\caption{Signals $pp\rightarrow t_{4}t_{4}\rightarrow 6t$,  $pp\rightarrow t_{4}t_{4}\rightarrow 4b2t$ and background fiducial cross sections in units of fb in the 4b analysis. For each $t_{4}$ mass hypothesis, the optimized cuts are listed in the last row for integrated luminosity $L=3\text{ ab}^{-1}$. The signal efficiency of the cuts $(\epsilon)$ is also presented. All results are presented using the 1b/2b-tagging working point $(\epsilon_{1b}, \bar{\epsilon}_{2b})=(0.8,1/3)$.} 
\label{tab:4bAnalysis_t4}
\end{center}
\end{table}

\newpage

As anticipated the expected bounds weaken as the number of top-jets in the final state increases. If decays to multiple heavy Higgs bosons are simultaneously open, the upper limit on the total $b_4$ ($t_4$) branching ratio lies in between the solid and dashed lines but depends on the relative size of branching ratios into $H$, $A$ and $H^\pm$. As an example, the dot-dashed lines show the cases in which the branching ratios into the three heavy Higgses, $H, A, H^{\pm}$, follow the typical patterns for singlet-like $b_{4}$ and $t_{4}$, $1/4:1/4:1/2$, assuming degenerate masses $m_{A,H,H^{\pm}}=1$ TeV. In this case, we also get contributions from processes in which the two vectorlike quarks decay to different Higgses. In particular this allows for the final states $4b2t$ and $4t2b$ from decays of $b_4$ and $t_4$, respectively. This is especially relevant for $H^\pm$ decays in $b_4$ cascades because the $4b2t$ final state is superior to the $4t2b$ one in the context of the all-hadronic analysis we are considering.

\begin{figure}[ht]
\centering
\includegraphics[scale=0.35]{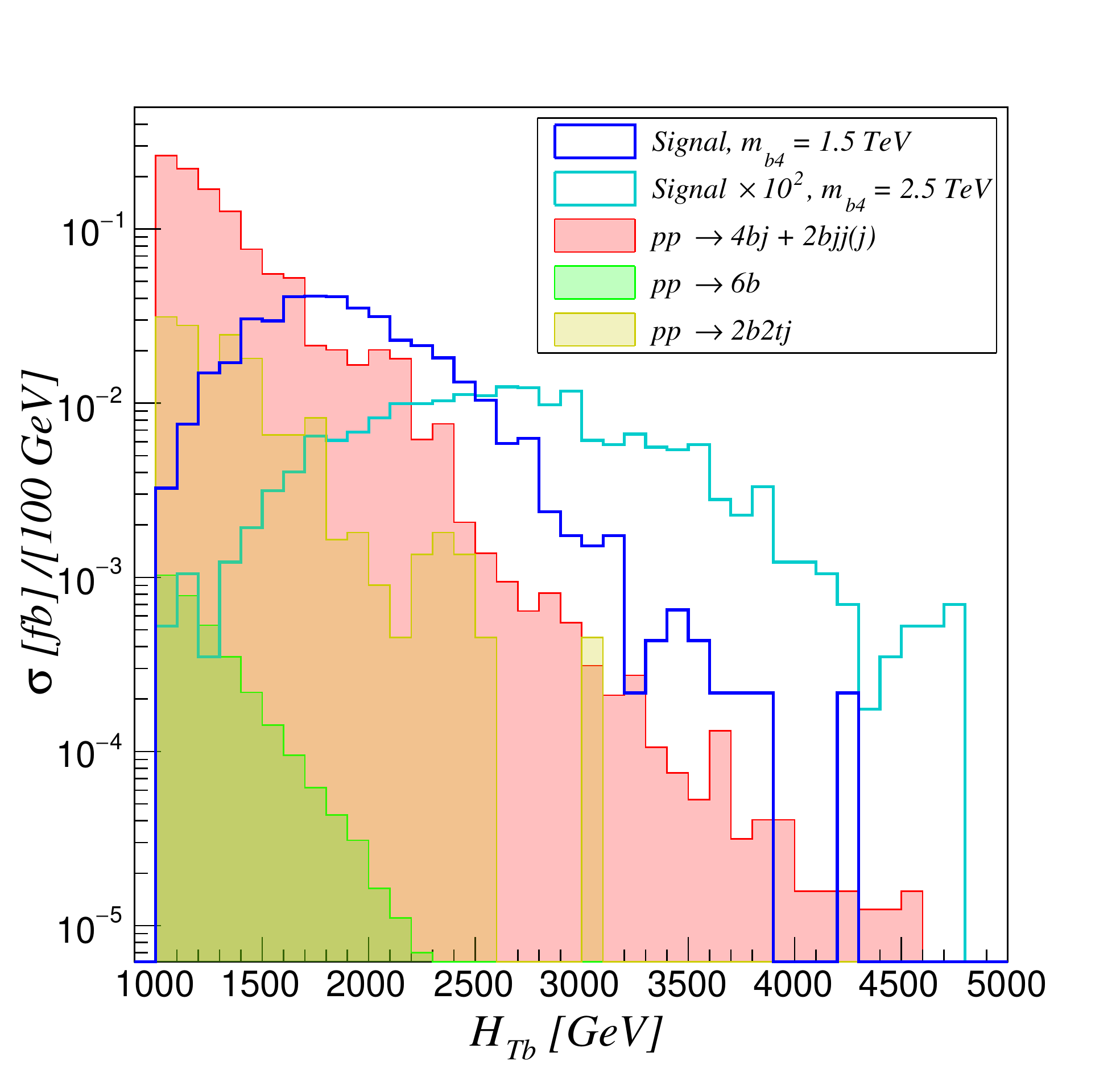} 
\includegraphics[scale=0.35]{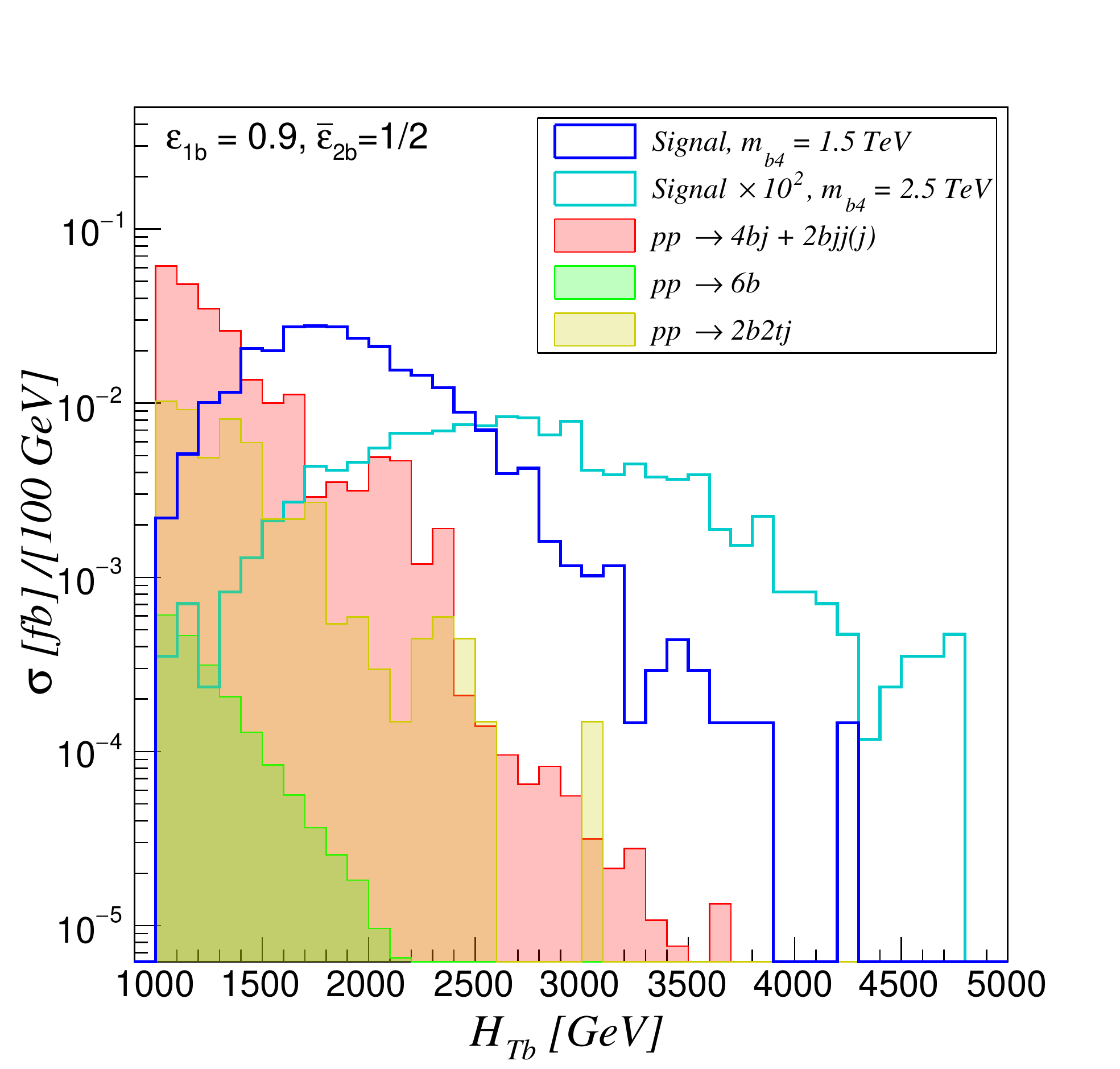} 
\caption{{\bf Left:} $H_{Tb}$ distribution of the $6b$ signal and backgrounds in the 5b analysis.  We show the signal distribution for $m_{b_4}=1.5$ and $2.5$ TeV in blue and cyan respectively. We show the dominant backgrounds, $4bj + 2bjj + 2bjjj$ in the red shaded region. We show the subdominant backgrounds, $6b$ and $2b2tj$, in the green and yellow shaded regions respectively. {\bf Right:}  $H_{Tb}$ distribution of the $6b$ signal and backgrounds in the 5b analysis after incorporating the 1b/2b-tagging strategy assuming $(\epsilon_{1b} ,\bar\epsilon_{2b}) = (0.9,1/2)$ efficiencies.}
\label{fig:5b_HTb}
\end{figure}

\subsection{5b analysis}
\label{sec:5b}

\begin{figure}[t]
\centering
	\includegraphics[scale=0.85]{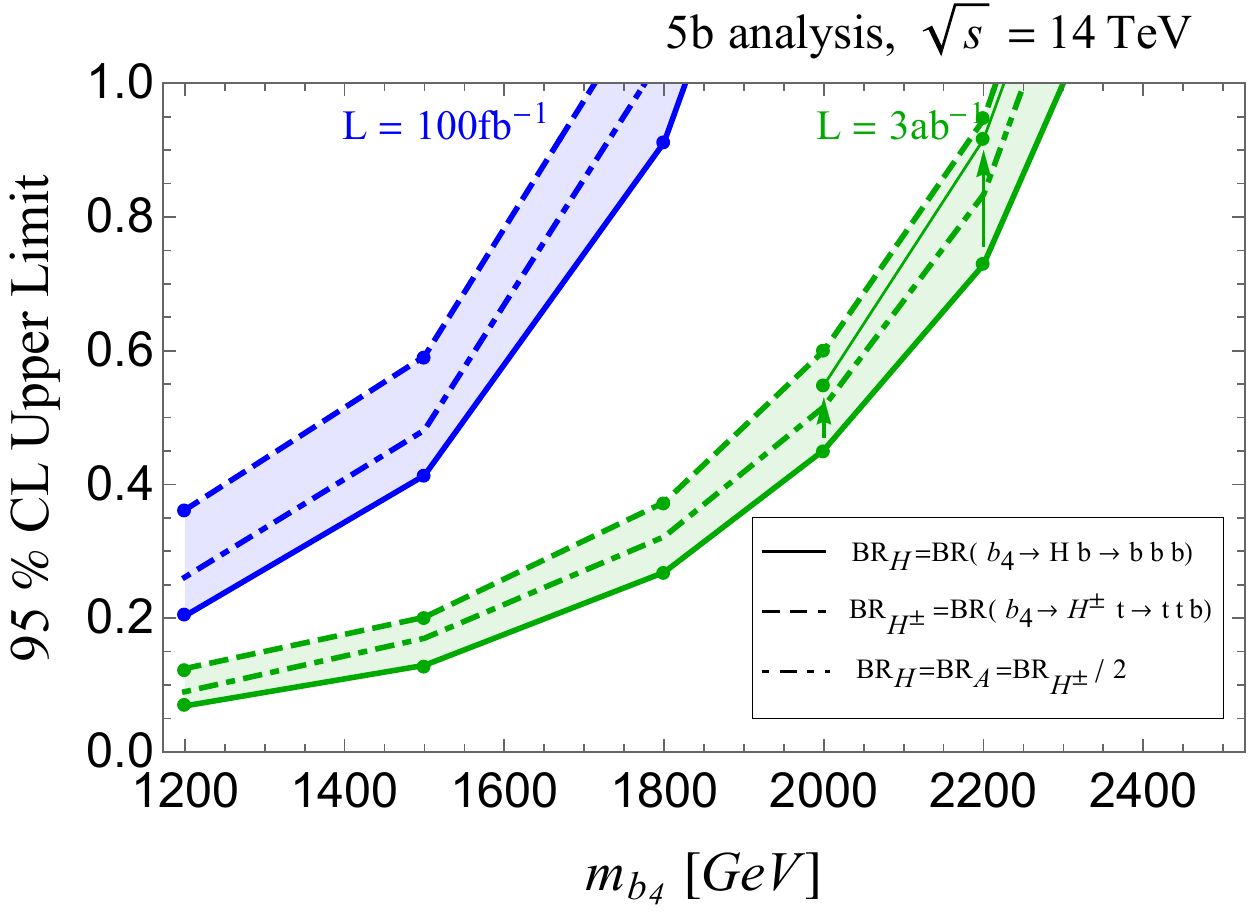} 
\caption{Expected $95\%$ CL upper limits on the $b_4$ branching ratio into neutral or charged Higgses in the 5b analysis. Styling of all curves matches that in figure~\ref{fig:4b_b4_bounds}.}
\label{fig:5b_b4_bounds}
\end{figure}

The dominant irreducible backgrounds to an analysis based on five high-$p_T$ $b$-tagged jets are $6b$, $6t$, $4b2t$ and $4t2b$. On top of this we have contributions from ($4b$, $4t$, $2b2t $) + jet and ($2b$,$2t$) + 3 jets where the jets are mistagged. Due to the sizable mistag rate coupled with a large diagrammatic multiplicity, the latter sources of backgrounds are by far dominant. The situation can be ameliorated by adopting the additional 1b/2b tagging strategy. 

As mentioned above, in the 5b analysis we are guaranteed to be able to reconstruct one full vectorlike quark decay. For each choice of the heavy Higgs and vectorlike quark masses we additionally optimize the significance by requesting that the invariant masses of at least one pair and one triplet of $b$-jets (obtained by adding one of the other jets to the first pair) reconstruct the signal decay chain. For signal with larger masses we find optimal cuts with $|m_{bbb}-m_{b_4}| < 1$ TeV and $|m_{bb} - m_{H}| < 400$ GeV, where the cut on the triplet invariant mass slightly decreases for smaller $m_{b_4}$.

\begin{figure}[h]
\centering
	\includegraphics[scale=0.85]{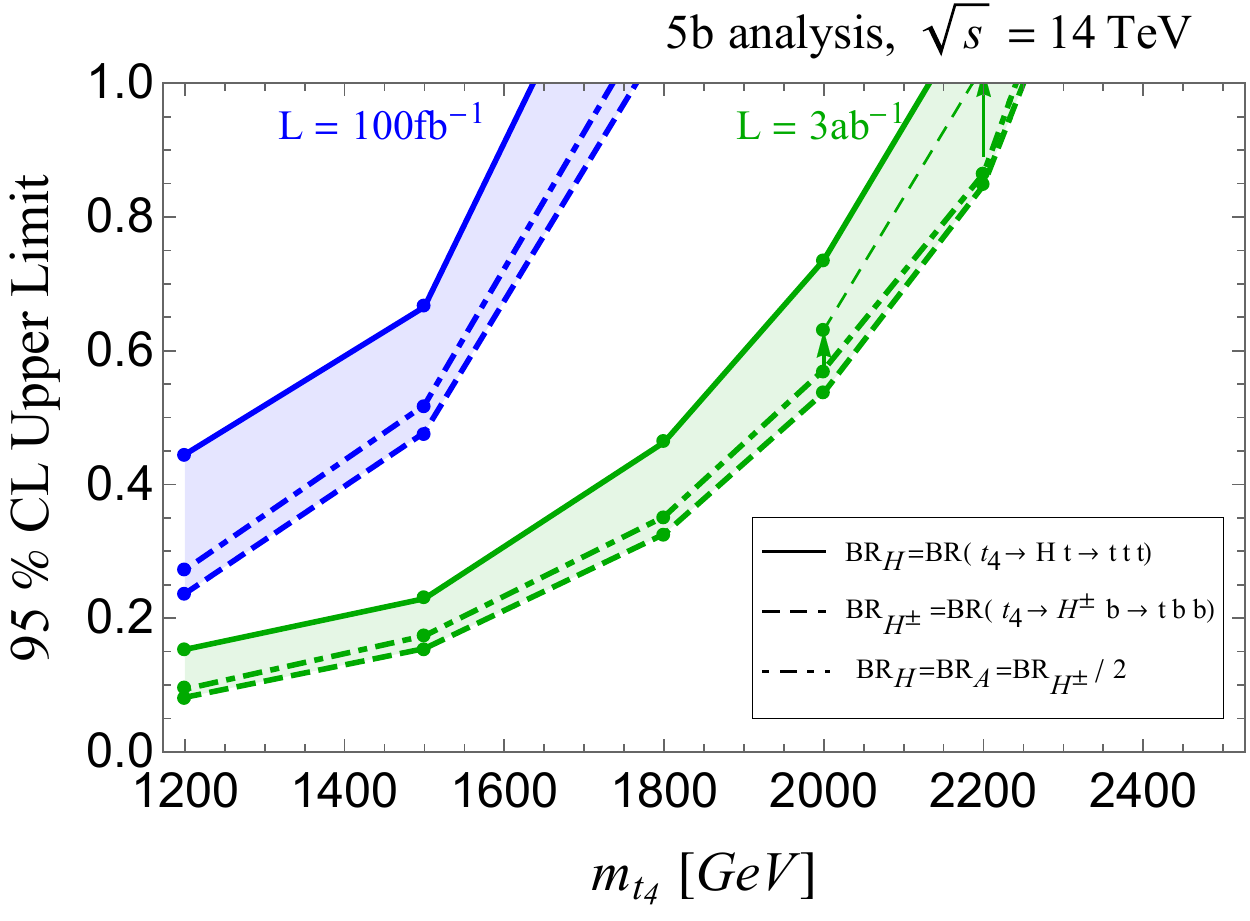} 
\caption{Expected $95\%$ CL upper limits on the $t_4$ branching ratio into neutral or charged Higgses in the 5b analysis. Styling of all curves matches that in figure~\ref{fig:4b_t4_bounds}.}
\label{fig:5b_t4_bounds}
\end{figure}

In figure~\ref{fig:5b_HTb} we show the $H_{Tb}$ distributions of signal and backgrounds. In tables~\ref{tab:5bAnalysis_b4} and \ref{tab:5bAnalysis_t4} we present the result of the $H_{Tb}^{\rm cut}$ optimization. The upper bounds on the total branching ratios of vectorlike quarks into heavy Higgses are presented in figures~\ref{fig:5b_b4_bounds} and \ref{fig:5b_t4_bounds}. The discussion of the tables and the bounds parallels the one presented in the previous section.

\begin{table}[h]
\begin{center}
\resizebox{!}{1.5in}{%
 \begin{tabular}{|c || c | c | c | c | c |} 
 \hline
 $m_{b_4}$  & 1200 & 1500 & 1800 & 2000 & 2200\\ [0.5ex] 
\hline\hline
$6b$ signal & $1.02$&$2.20\times 10^{-1}$ &$4.01\times 10^{-2}$ &$1.42\times 10^{-2}$ & $5.43\times 10^{-3}$\\
\hline
$\epsilon_{signal}$ & 4.5 \% & 7.5 \% & 5.8 \% & 6.0 \% & 5.6\%\\
\hline
2bjj(j)& $2.96\times 10^{-2}$ & $1.32\times 10^{-2}$ & $5.98\times 10^{-3}$ & $5.98\times 10^{-3}$ & $5.98\times 10^{-3}$ \\
\hline
4bj& $3.10\times 10^{-3}$ & $2.32\times 10^{-3}$&$1.55\times 10^{-3}$ &$1.55\times 10^{-3}$ & $1.55\times 10^{-3}$\\
\hline
2t2bj & $4.43\times 10^{-4}$& $4.15\times 10^{-4}$ & $3.60\times 10^{-4}$ & $3.60\times 10^{-4}$ & $3.60\times 10^{-4}$\\
\hline
6b& $6.58\times 10^{-5}$& $3.69\times 10^{-5}$ & $1.90\times 10^{-5}$ & $1.90\times 10^{-5}$ & $1.90\times 10^{-5}$\\
\hline
$H^{cut}_{T_b}$& $1250$ & $1500$ & $1750$ & $1750$& 1750\\
\hline\hline
$4t2b$ signal & $3.74\times 10^{-1}$&$1.19\times 10^{-1}$ &$2.65\times 10^{-2}$ &$8.02\times 10^{-3}$ & $4.34\times 10^{-3}$\\
\hline
$\epsilon_{signal}$ & 4.9 \% & 5.9 \% & 9.1 \% & 11 \% & 5.1\%\\
\hline
2bjj(j)& $5.68\times 10^{-2}$ & $2.96\times 10^{-2}$ & $1.33\times 10^{-2}$ & $5.99\times 10^{-3}$ & $5.99\times 10^{-3}$ \\
\hline
4bj& $3.10\times 10^{-3}$ & $3.10\times 10^{-3}$&$2.32\times 10^{-3}$ &$1.55\times 10^{-3}$ & $1.55\times 10^{-3}$\\
\hline
2t2bj & $4.43\times 10^{-4}$& $4.43\times 10^{-4}$ & $4.15\times 10^{-4}$& $3.60\times 10^{-4}$ & $3.60\times 10^{-4}$\\
\hline
6b& $7.20\times 10^{-5}$& $6.58\times 10^{-5}$ & $3.68\times 10^{-5}$ & $1.90\times 10^{-5}$ & $1.90\times 10^{-5}$\\
\hline
$H^{cut}_{T_b}$& $1000$ & $1250$ & $1500$ & $1750$& 1750\\
\hline
\end{tabular}}
\caption{Signals $pp\rightarrow b_{4}b_{4}\rightarrow 6b$,  $pp\rightarrow b_{4}b_{4}\rightarrow 4t2b$ and background fiducial cross sections in units of fb in the 5b analysis. For each $b_{4}$ mass hypothesis, the optimized cuts are listed in the last row for integrated luminosity $L=3\text{ ab}^{-1}$. The signal efficiency of the cuts $(\epsilon)$ is also presented. All results are presented using the 1b/2b-tagging working point $(\epsilon_{1b}, \bar{\epsilon}_{2b})=(0.9,0.5)$.} 
\label{tab:5bAnalysis_b4}
\end{center}
\end{table}



\begin{table}[h]
\begin{center}
\resizebox{!}{1.5in}{%
 \begin{tabular}{|c || c | c | c | c | c |} 
 \hline
 $m_{t_4}$  & 1200 & 1500 & 1800 & 2000 & 2200\\ [0.5ex] 
\hline\hline
$6t$ signal & $2.48\times 10^{-1}$&$1.10\times 10^{-1}$ &$1.72\times 10^{-2}$ &$6.81\times 10^{-3}$ & $2.23\times 10^{-3}$\\
\hline
$\epsilon_{signal}$ & 2.0 \% & 4.6 \% & 3.4 \% & 3.5 \% & 2.9\%\\
\hline
2bjj(j)& $5.67\times 10^{-2}$ & $5.67\times 10^{-2}$ & $1.33\times 10^{-2}$ & $1.33\times 10^{-2}$ & $5.98\times 10^{-3}$ \\
\hline
4bj& $3.10\times 10^{-3}$ & $3.10\times 10^{-3}$ & $2.32\times 10^{-3}$ & $2.32\times 10^{-3}$  & $1.55\times 10^{-3}$\\
\hline
2t2bj & $4.43\times 10^{-4}$& $4.15\times 10^{-4}$ & $4.615\times 10^{-4}$ & $4.615\times 10^{-4}$ & $3.60\times 10^{-4}$\\
\hline
6b& $7.20\times 10^{-5}$ & $7.20\times 10^{-5}$ & $3.69\times 10^{-5}$ & $3.69\times 10^{-5}$ & $1.90\times 10^{-5}$\\
\hline
$H^{cut}_{T_b}$& $1000$ & $1000$ & $1500$ & $1500$& 1750\\
\hline\hline
$4b2t$ signal & $8.74\times 10^{-1}$&$1.55\times 10^{-1}$ &$2.72\times 10^{-2}$ &$9.99\times 10^{-3}$ & $4.02\times 10^{-3}$\\
\hline
$\epsilon_{signal}$ & 6.9 \% & 6.5 \% & 5.3 \% & 5.1 \% & 5.3\%\\
\hline
2bjj(j)& $5.68\times 10^{-2}$ & $1.32\times 10^{-2}$ & $5.99\times 10^{-3}$ & $5.99\times 10^{-3}$ & $5.99\times 10^{-3}$ \\
\hline
4bj& $3.10\times 10^{-3}$ & $2.32\times 10^{-3}$&$1.55\times 10^{-3}$ &$1.55\times 10^{-3}$ & $1.55\times 10^{-3}$\\
\hline
2t2bj & $4.43\times 10^{-4}$& $4.15\times 10^{-4}$ & $3.60\times 10^{-4}$& $3.60\times 10^{-4}$ & $3.60\times 10^{-4}$\\
\hline
6b& $7.20\times 10^{-5}$& $3.69\times 10^{-5}$ & $1.90\times 10^{-5}$ & $1.90\times 10^{-5}$ & $1.90\times 10^{-5}$\\
\hline
$H^{cut}_{T_b}$& $1000$ & $1500$ & $1750$ & $1750$& 1750\\
\hline
\end{tabular}}
\caption{Signals $pp\rightarrow t_{4}t_{4}\rightarrow 6t$,  $pp\rightarrow t_{4}t_{4}\rightarrow 4b2t$ and background fiducial cross sections in units of fb in the 5b analysis. For each $t_{4}$ mass hypothesis, the optimized cuts are listed in the last row for integrated luminosity $L=3\text{ ab}^{-1}$. The signal efficiency of the cuts $(\epsilon)$ is also presented. All results are presented using the 1b/2b-tagging working point $(\epsilon_{1b}, \bar{\epsilon}_{2b})=(0.9,0.5)$.} 
\label{tab:5bAnalysis_t4}
\end{center}
\end{table}

\newpage

\subsection{Discussion and summary of results}
In the 5b analysis we find that the LHC with 3 ab$^{-1}$ of integrated luminosity is sensitive to the $b_4 \to H (A) b \to bbb$ decay for masses of $b_4$ up to about 2.2 TeV and masses of neutral Higgs bosons up to about 2 TeV (taking into account that the $H\to b\bar b$ branching ratio is slightly less than 100\% due to the presence of additional $H$ decays). The same analysis, applied for $t_4 \to H^\pm b \to tbb$, leads to a similar reach: the masses of $t_4$ up to about 2.2 TeV and the charged Higgs boson up to almost 2 TeV can be explored at the LHC. With already existing data accumulated at 13 TeV LHC corresponding to 139 fb$^{-1}$ of integrated luminosity we estimate that the presented search strategy is sensitive to vectorlike quark masses up to about 1.8 TeV and charged and neutral Higgs bosons up to about 1.6 TeV.

The 4b analysis has a slightly lower reach, but the lower $b$-jets multiplicity allowed us to perform an in depth study of multi-jet QCD backgrounds and to demonstrate that they can be brought under control. In particular, we have studied the breakdown of $4b+j$ and $2b +jj(j)$ backgrounds and their contributions to the $4b$ final state after shower effects. We found that a significant effect originates from jets in which a gluon splits into $b\bar{b}$ pairs. However, separation of genuine $b$-jets and 2b-jets can be used reduce the contribution of QCD mulit-jet backgrounds by about an order of magnitude (see figure~\ref{fig:4b_HTb}) with only a mild effect on a genuine $4b$ signal. This makes the analysis robust against backgrounds with high jet multiplicity, while still remaining conceptually simple.

\begin{figure}[ht]
\centering
	\includegraphics[width=0.49 \linewidth]{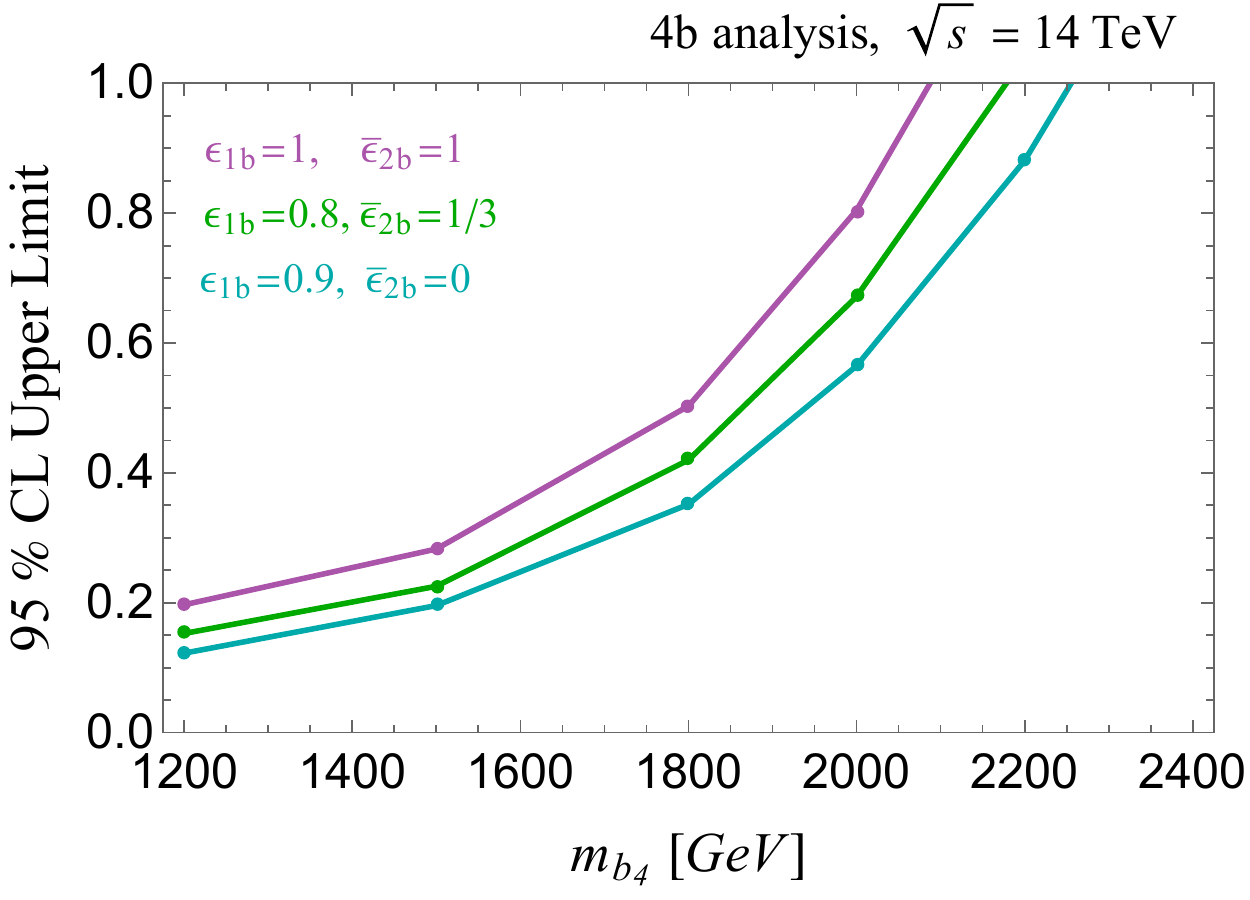} 
	\includegraphics[width=0.49 \linewidth]{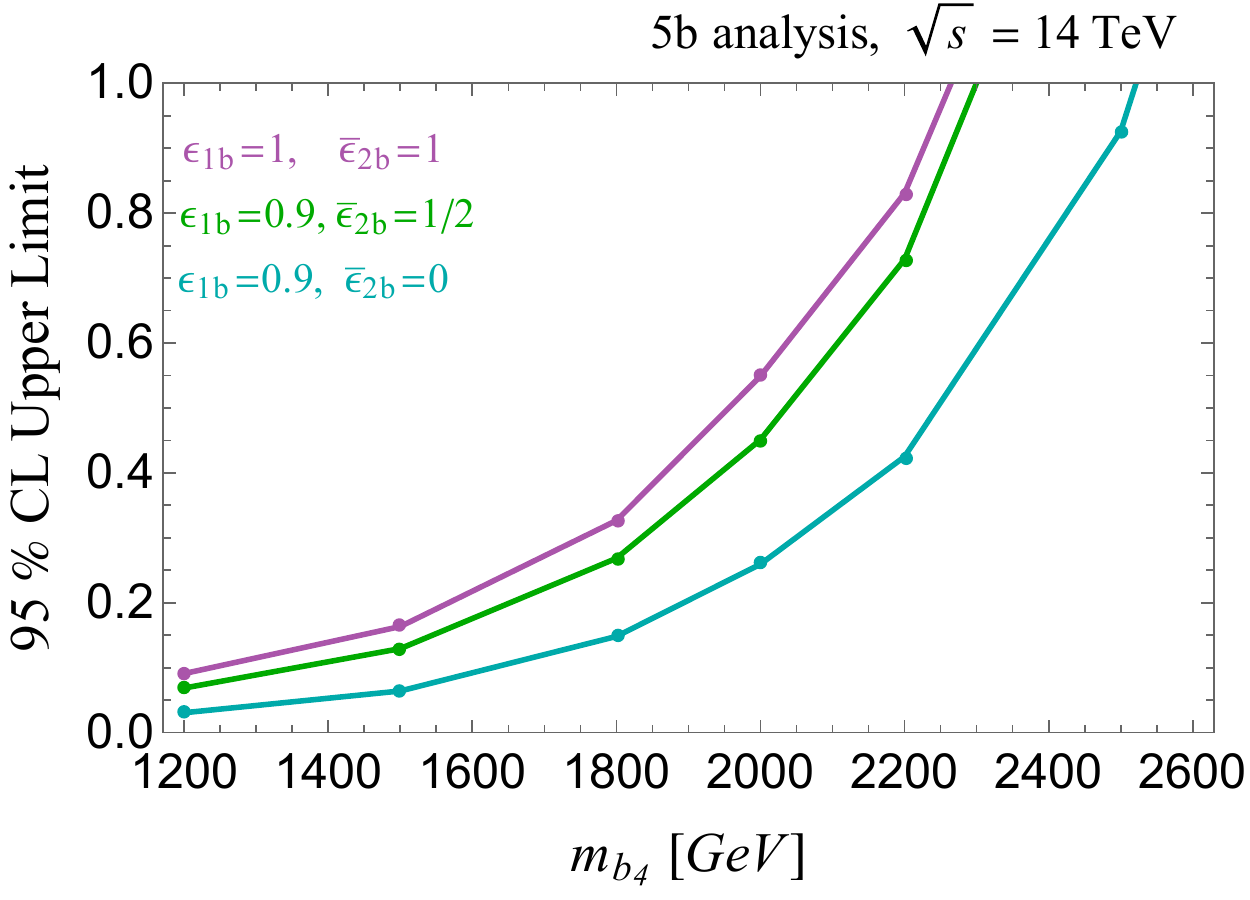} 
\caption{{\bf Left}: Impact of 1b/2b tagging efficiencies on the 4b analysis. In purple, we show the expected $95\%$ CL upper limit on the $b_4\rightarrow H b\rightarrow bbb$ branching ratio with $m_{H} = 1$ TeV assuming no discrimination between genuine b-jets and bb-jets. In green we show the optimal tagging efficiencies used in the main results. We show the effect of perfect discrimination of 2b-jets in Cyan. {\bf Right:} Corresponding study of tagging efficiencies in the 5b analysis.}
\label{fig:e1b_e2b_bounds}
\end{figure}

In both analyses, we incorporate a 1b/2b tagging efficiency that can be obtained in a simple cut-based analysis \cite{Goncalves:2015prv}. Though, improvements in the reach of vectorlike quark and heavy Higgs masses can be achieved with stronger rejection of 2b-jets. In figure~\ref{fig:e1b_e2b_bounds} we show the impact of the 1b/2b tagging strategy on the reach of the 6$b$ signal. Purple and green curves are the results obtained without and with the adoption of the 1b/2b discrimination. The blue curves show the limiting case of perfect 2b rejection. Note that, for the 5b analysis, improvements in the 1b/2b separation have the potential to considerably improve the bounds. In any case, rejection of 2b-jets is critical for our confidence in the analyses and efficient rejection of multi-jet QCD background.

\section{Conclusions}
\label{sec:conclusions}

We discussed search strategies for cascade decays of pair-produced vectorlike quarks through charged or neutral Higgs bosons that lead to 6 top or bottom quarks in final states. 
In extensions of  the  type-II  2HDM with vectorlike quarks, such cascade decays can easily dominate. Depending on model parameters and especially on the hierarchies in the masses of new quarks and Higgs bosons, many decay chains are possible, and several of them can be simultaneously sizable. Among them, $t_4 \to H^\pm b \to tbb$ and $b_4 \to H (A) b \to bbb$ are especially well motivated, since the corresponding branching ratios can be close to 100\% for any medium to large $\tan\beta$, even if all the decay modes are kinematically open. The suggested search strategies, focusing on the $6b$ final state, are tailored for $b_4 \to bH \to bbb$ process. However, we also find them very effective for  $t_4 \to H^\pm b \to tbb$  and  other  possible cascade decays of $t_4$ and $b_4$.

While the signals we consider are subject to large reducible backgrounds originating from QCD multi-jets, we find that a significant portion of these backgrounds can be rejected by separating genuine $b$-jets from those emerging from parton shower effects. This aspect of our analysis gives us confidence that our results are robust against QCD backgrounds with high jet multiplicity which otherwise would pose a formidable challenge to properly estimate.

Among the main results we find that the Large Hadron Collider with 3 ab$^{-1}$ of integrated luminosity is sensitive to $b_4 \to H (A) b \to bbb$ for masses of $b_4$ up to about 2.2 TeV and masses of neutral Higgs bosons up to about 2 TeV. The same analysis, applied for $t_4 \to H^\pm b \to tbb$, leads to a similar reach: the masses of $t_4$ up to about 2.2 TeV and the charged Higgs boson up to almost 2 TeV can be explored at the LHC. The sensitivity of the analysis based on the $6b$ final state gradually loosens as the number of top-jets increases. However, we find that even for the $6t$ final state from pair produced $t_4$ the sensitivity still extends to $t_4$ masses of about 2 TeV. 

With already existing data accumulated at 13 TeV LHC corresponding to 139 fb$^{-1}$ of integrated luminosity we estimate that the presented search strategy is sensitive to vectorlike quark masses up to about 1.8 TeV and charged and neutral Higgs bosons up to about 1.6 TeV.

The reach of the presented search strategy significantly exceeds the reach of separate searches for vectorlike quarks and charged and neutral Higgs bosons.  
Especially the sensitivity to the charged Higgs boson, extending to about 2 TeV, stands out when compared to models without vectorlike matter. 
Thus searching for bottom-rich events presents a unique opportunity to simultaneously discover vectorlike quarks and heavy charged and neutral Higgs bosons at the LHC.

\vspace{0.5cm}
\noindent
{\bf Acknowledgments:} This work was supported in part by the U.S. Department of Energy under grant number {DE}-SC0010120. The work of NM is supported by the U.S. Department of Energy, Office of Science, Office of Work- force Development for Teachers and Scientists, Office of Science Graduate Student Research (SCGSR) program. The SCGSR program is administered by the Oak Ridge Institute for Science and Education (ORISE) for the DOE. ORISE is managed by ORAU under contract number de-sc0014664. The work of SS is supported by the National Research Foundation of Korea (NRF-2017R1D1A1B03032076 and NRF-2020R1I1A3072747).
This work was performed at the Aspen Center for Physics, which is supported by National Science Foundation grant PHY-1607611.
SS would like to express a special thanks to the Mainz Institute for Theoretical Physics (MITP) of the Cluster of Excellence PRISMA+ (Project ID 39083149) for its hospitality and support.

\appendix

\section{1b/2b jet discrimination}
\label{app:rescaling}
In order to estimate the impact of the 1b/2b tagging, we use simple combinatorics to calculate the additional efficiencies which we impose on signal and backgrounds.

We start by giving the efficiencies for tagging at least $n$ $b$-jets out of $m$ jets containing a $b$-hadron:
\begin{align}
f_{n \in n} (\epsilon_b) &= \epsilon_b^n \; ,\\
f_{4 \in 5} (\epsilon_b) &= \epsilon_b^5 + 5 \epsilon_b^4 (1-\epsilon_b) \; , \\
f_{4 \in 6} (\epsilon_b) &= \epsilon_b^6 + 6 \epsilon_b^5 (1-\epsilon_b) + 15 \epsilon_b^4 (1-\epsilon_b)^2 \; ,\\
f_{5 \in 6} (\epsilon_b) &= \epsilon_b^6 + 6 \epsilon_b^5 (1-\epsilon_b)  \; ,
\end{align} 
where $\epsilon_{b}$ is the $b$-tagging efficiency. We will also need the efficiency for tagging at least 4 or 5 $1$b-jets out of 6 jets containing a $b$-hadron:
\begin{align}
f_{4 \in 6} (\epsilon_b,\epsilon_{1b}) &= \epsilon_b^6 f_{4 \in 6} (\epsilon_{1b})  +  6 \epsilon_b^5 (1-\epsilon_b) f_{4 \in 5} (\epsilon_{1b})+ 15 \epsilon_b^4 (1-\epsilon_b)^2 f_{4 \in 4} (\epsilon_{1b}) \; ,\\
f_{5 \in 6} (\epsilon_b,\epsilon_{1b}) &= \epsilon_b^6 f_{5 \in 6} (\epsilon_{1b})  +  6 \epsilon_b^5 (1-\epsilon_b) f_{5 \in 5} (\epsilon_{1b})\; .
\end{align}
For the 4$b$ analysis, the additional efficiencies due to the 1b/2b discrimination can be estimated as:
\begin{align}
\sigma ({\rm signal}) &\to f_{4 \in 6} (\epsilon_b,\epsilon_{1b}) / f_{4 \in 6} (\epsilon_b) \; ,\\
\sigma (2bjj) &\to   \epsilon_{1b}^2  \bar\epsilon_{2b}^2\; , \\
\sigma (4b,2b2t,4t) &\to   \epsilon_{1b}^4 \; .
\end{align}
For the 5$b$ analysis, the additional efficiencies due to the 1b/2b discrimination can be estimated as:
\begin{align}
\sigma ({\rm signal}) &\to f_{5 \in 6} (\epsilon_b,\epsilon_{1b}) / f_{5 \in 6} (\epsilon_b) \; ,\\
\sigma (2bjjj) &\to   \epsilon_{1b}^2  \bar\epsilon_{2b}^3\; , \\
\sigma (4bj,2b2tj,4tj) &\to   \epsilon_{1b}^4 \bar\epsilon_{2b}\; .
\end{align}

\section{Poisson statistics}
\label{app:poisson}
For completeness we present the statistical framework we adopt to extract the bounds. Following ref.~\cite{Tanabashi:2018oca}, the upper limit (at confidence level of $1-\alpha$) of the number of signal events that can be extracted from observing $n$ events over an expected number of background events $b$ is:
\begin{align}
s_{\rm up} (\alpha, n, b) &= \frac{1}{2} F_{\chi^2}^{-1} [p, 2 (n+1)] -b \\
p &= 1- \alpha (1 - F_{\chi^2} [2b, 2(n+1)]) ,
\end{align}
where $F_{\chi^2}$ is the $\chi^2$ cumulative distribution. It is easy to show that, assuming no signal and Poisson distributed background, the median of the upper limits calculated over a sample of the background distribution is given by:
\begin{align}
\bar s_{\rm up} (\alpha, b) = s_{\rm up} (\alpha, b, b) \; .
\end{align}
Note that for $\alpha = 0.95$ and at large $b$, one finds $\bar s_{\rm up} (0.95,b) \simeq 2 \sqrt{ b + 2 \sqrt{b}}$ as expected. The upper limit on the signal cross section is then
\begin{align}
\sigma_{\rm sig}^{\rm up} = \frac{ \bar s_{\rm up} (\alpha, \sigma_B {\cal L})}{{\cal L} \cdot \varepsilon_{\rm sig}},
\end{align}
where ${\cal L}$ is the integrated luminosity and $\varepsilon_{\rm sig}$ is the signal efficiency of the adopted cuts. Finally, the signal cross sections that we consider are proportional to the square of the vectorlike quark branching ratio for which we calculate the expected upper limits (e.g. $\sigma (pp \to b_4 \bar b_4) \; [{\rm BR} (b_4 \to b \bar b b) ]^2 $).

\bibliography{6b}{}
\bibliographystyle{JHEP} 

\end{document}